%% file: kpd1930astroPH.tex
\documentstyle[longtable,epsfig,epsf,onecolumn]{mn}

\title[WET Observations of KPD~1930+2752]
{Whole Earth Telescope 
Observations of the subdwarf B star KPD~1930+2752: A rich, short period 
pulsator in a close binary.}
\author[M. D. Reed et al.]{M. D. Reed,$^{1}$\thanks{E-mail:
MikeReed@missouristate.edu}\thanks{Visiting Astronomer, Kitt Peak National Observatory, National Optical Astronomy Observatory, which is operated by the Association of Universities for Research in Astronomy (AURA) under cooperative agreement with the National Science Foundation.}, S.L. Harms,$^{1}$ S. Poindexter,$^{1,2}$ A.-Y.
Zhou,$^{1,3}$ J.R. Eggen$^{1}$,\cr M.A. Morris$^{1,4}$, A.C. Quint$^1$,
S. McDaniel,$^{1}$ A. Baran,$^5$, N. Dolez,$^6$
S. D. Kawaler,$^7$\cr  D. W. Kurtz,$^8$
 P. Moskalik,$^9$ R. Riddle,$^{7,10}$ 
S. Zola,$^{5,11}$
R. H. \O stensen,$^{12,13}$ \cr J.-E. Solheim,$^{14}$ S.O. Kepler,$^{15}$
A.~F.~M. Costa,$^{15b}$ 
J. L. Provencal,$^{16}$
F. Mullally,$^{17}$\cr D. W. Winget,$^{17}$ M. Vuckovic,$^{7,13}$ 
R. 
Crowe,$^{18}$ D. Terry, $^{18}$ R. Avila,$^{18,19}$ B. Berkey,$^{18,20}$ \cr
S. Stewart,$^{18}$ J. Bodnarik,$^{18,21}$ D. Bolton,$^{18}$ P.-M. Binder,$^{18}$
 K. Sekiguchi,$^{22}$\cr D. J. Sullivan,$^{23}$ 
S.-L. Kim,$^{24}$ W.-P. Chen,$^{25}$ C.-W. Chen,$^{25}$
H.-C. Lin,$^{25}$
 X.-J. Jian,$^{26}$\cr H. Wu,$^{26}$ J.-P. Gou,$^{26}$ Z. Liu,$^{26}$ 
E. Leibowitz,$^{27}$ Y. Lipkin,$^{27}$
C. Akan,$^{28}$ O. Cakirli,$^{28}$\cr R. Janulis,$^{29}$  R.
Pretorius,$^{30}$ W. Ogloza,$^5$ G.
Stachowski,$^5$ 
M. Paparo,$^{31}$ R. Szabo,$^{31}$\cr Z. Csubry,$^{31}$ D. Zsuffa,$^{31}$ 
R. Silvotti,$^{32}$ S.
Marinoni,$^{33,34}$ I. Bruni,$^{34}$ G. Vauclair,$^{6}$ \cr
M. Chevreton,$^{35}$ 
J.M. Matthews$^{36}$, C. Cameron$^{36}$, and H. Pablo$^7$ \\
$^{1}$Department of Physics, Astronomy, and Materials Science,
Missouri State University, Springfield, MO, 65897, USA\\
$^{2}$ Department of Astronomy, The Ohio State University,
140 W. 18th Ave., Columbus, OH 43210 USA.\\
$^3$ National Astronomical Observatories, Chinese Academy of Sciencies, Beijing 100012, China\\
$^{4}$ School of Earth and Space Exploration, Arizona State University, 
P.O. Box 871404, Tempe, Arizona 85287-1404 \\
$^{5}$Mt. Suhora Observatory of the Pedagogical University, ul. 
Podchor\c{a}zych 2, PL-30-084 Kracow, Poland\\
$^{6}$Laboratoire d'Astrophysique de Toulouse-Tarbes, Universit\'e de Toulouse, CNRS, 14 Av. Ed. Belin, 31400 Toulouse, France\\
$^{7}$ Department of Physics and Astronomy, Iowa State University, Ames, 
IA 50011, USA \\
$^{8}$Centre for Astrophysics, University of Central Lancashire, 
Preston PR1 2HE  \\
$^{9}$Nicolaus Copernicus Astronomical Center, ul. Bartycka 18, 00-716 Warsaw, 
Poland  \\
$^{10}$ California Institute of Technology, 1200 E. California Blvd.  MC 11-17,
Pasadena, CA 91125 USA \\
$^{11}$ Astronomical Observatory, Jagiellonian University, ul. 
Orla 171, 30-244 Cracow, Poland \\
$^{12}$ Isaac Newton Group of Telescopes, 37800 Santa Cruz de La Palma, Spain \\
$^{13}$ Instituut voor Sterrenkunde, Katholieke Universiteit Leuven, Celestijnenlaan 200 D, 3001 Leuven, Belgium\\
$^{14}$ Institutt for Teoretisk Astrofysikk, Universitetet i Oslo, 0212, Blindern-Oslo, Norway\\
$^{15}$ Instituto de Fisica, UFRGS, CP 15051, 91501-970 Porto Alegre, RS, Brazil\\
$^{15b}$Universidade do Estado de Santa Catarina, CAV, CP 88520-000 Lages, SC, Brazil\\
$^{16}$  Mount Cuba Observatory and University of Delaware, Newark, DE 19716\\
$^{17}$ McDonald Observatory, and Department of Astronomy, University of Texas, Austin, TX 78712, USA \\
$^{18}$ Department of Physics and Astronomy, University of Hawaii - Hilo, 
200 West Kawili Street, Hilo, Hawaii, 96720-4091, USA \\
$^{19}$ Department of Astronomy, New Mexico State University, Box 30001, Las Cruces, NM 88003, USA\\
$^{20}$ W. M. Keck Observatory, Kamuela, HI 96743, USA \\
$^{21}$ Gemini Observatory, 670 North A'ohoku Place, Hilo, HI 96720, USA\\
$^{22}$ Subaru Observatory, National Astronomical Observatory of Japan, 
650 North A'ohoku Place, Hilo, HI 96720, USA \\
$^{23}$ School of Chemical and Physical Sciences, Victoria University of 
Wellington, PO Box 600, Wellington, New Zealand \\
$^{24}$  Korea Astronomy Observatory, Daejeon 305-348, Korea \\
$^{25}$ Graduate Institute of Astronomy, National Central University, 
Chung-Li, Taiwan \\
$^{26}$  National Astronomical Observatories and Joint Laboratory of 
Optical Astronomy, Chinese Academy of Sciences, Beijing, 100012, China\\
$^{27}$  Wise Observatory, Tel-Aviv University, Israel\\
$^{28}$  Ege University Observatory, TR-35100 Bornova, \.{I}zmir, Turkey\\
$^{29}$  Institute of Theoretical Physics \& Astronomy of Vilnius University. A.
Go tauto St. 12, 01108 Vilnius, Lithuania\\
$^{30}$  South African Astronomical Observatory, PO Box 9, Observatory 7935, 
South Africa\\
$^{31}$ Konkoly Observatory, Box 67, H-1525 Budapest XII, Hungary \\
$^{32}$ INAF- Osservatorio Astronomico di Torino, strada dell'Osservatorio 20,
10025 Pino Torinese, Italy\\
$^{33}$  Universit\'{a}di Bologna, Dipartimento di Astronomia, via
Ranzani 1 40127, Bologna Italy and\\
Fundaci\'{o}n Galileo Galilei, INAF Rambla Jos\'{e} Ana Fern\'{a}ndez
P\'{e}rez, 7 38712 Bre\~{n}a Baja, TF, Spain\\
$^{34}$  INAF-Osservatorio Astronomico di Bologna, Via Ranzani 1, 40127 
Bologna, Italy \\
$^{35}$ LESIA, Observatoire de Paris-Meudon, Meudon, France\\
$^{36}$ Department of Physics and Astronomy, University of British 
Columbia, 6224 Agricultural Road, Vancouver, BC V6T1Z1, Canada \\
}
\begin{document}

\date{Accepted  Received }

\pagerange{\pageref{firstpage}--\pageref{lastpage}} \pubyear{2008}

\maketitle

\label{firstpage}

\begin{abstract}
KPD~1930+2752 is a short-period pulsating subdwarf B (sdB) star. It is
also an ellipsoidal variable with a known binary period just over
two hours. The companion is most likely a white dwarf and the total
mass of the system is close to the Chandresakhar limit. In this 
paper we report the results of Whole Earth Telescope (WET) photometric 
observations during 2003 and
a smaller multisite campaign from 2002. From 355 hours of WET data,
we detect  68 pulsation frequencies and suggest an additional 13 frequencies 
within a
crowded and complex temporal spectrum between 3065 and 6343~$\mu$Hz (periods
between 326 and 157~s). We examine pulsation properties including phase
and amplitude stability in an attempt to understand the nature of the pulsation
mechanism. 
We examine a stochastic mechanism by comparing amplitude variations
with simulated stochastic data.
We also use the binary nature of  KPD~1930+2752 for identifying pulsation
modes via multiplet structure and a tidally-induced pulsation geometry.
Our results indicate a complicated pulsation structure that includes
short-period ($\approx 16$~h) amplitude variability, rotationally split
modes, tidally-induced modes, and some pulsations which are geometrically
limited on the sdB star.

\end{abstract}


\section{Introduction}
Subdwarf B (sdB) stars are thought to have
 masses about 0.5M$_{\odot}$
with thin ($<10^{-2}$M$_{\odot}$) hydrogen shells and temperatures
from 22\,000 to 40\,000~K (Heber 1984; Saffer et al. 1994).
Shell hydrogen burning cannot be supported by such thin envelopes
 and it is likely they
proceed directly to the white dwarf
 cooling track without reaching the asymptotic
giant branch (Saffer \emph{et al.} 1994).

Pulsating sdB stars with periods of a few minutes
(officially V361~Hya, and also known as
EC~14026 or sdBV stars) were first observed by Kilkenny et al. (1997);
nearly simultaneous to their predicted existence by Charpinet et al.
(1996, 1997). The sdBV stars have pulsation periods ranging from
90 to 600 seconds with amplitudes typically near or below 1\%.
The pulsations are likely $p-$modes driven by the $\kappa$ mechanism 
due to a diffusive iron-group opacity
bump in the envelope (Charpinet et al. 1997, Jeffery \& Saio 2007).
Subdwarf B pulsators are typically found among the
hotter sdB stars, with T$_{\rm eff}\approx$34\,000~K and
$\log g\approx 5.8$. Reviews of this pulsation class include
Reed et al. (2007a) for an observational review of 23 resolved
class members and Charpinet et al. (2001) for a description of the pulsation
mechanism. Another class of pulsating sdB stars have periods longer
then 45 minutes, are likely $g-$mode pulsations and are designated
V1093~Her, but commonly known as  PG~1716 
stars (Green et al. 2003; Reed et al. 2004a). General reviews of sdB stars
and pulsators are Heber (2009) and \O stensen (2009).
Our target is a $p$-mode, sdBV-type pulsator.

KPD~1930+2752 (also V2214~Cyg and hereafter KPD~1930) was discovered to be a variable by
Bill\'eres et al. (2000; hereafter B00), who obtained data during
four nights
within one week. Their longest run was five hours, yet within this limited 
data set, they detected 45 separate
frequencies, which indicates that KPD~1930 is an interesting and complex
pulsator. A velocity study confirmed the $2^{\rm h}17^{\rm m}$ binary period
and, using the canonical sdB mass of 0.5~$M_{\odot}$, determined the
companion mass to be $0.97\pm 0.01M_{\odot}$ (Maxted, Marsh, \& North 2000). 
As the companion is not
observed either photometrically or spectroscopically, it is likely a
white dwarf, placing the mass of the system over the Chandrasekhar
limit. A study by Ergma, Fedorova, \& Yungelson (2001) suggested that the binary
will shed sufficient mass to avoid a type Ia supernova and will merge to
form a massive white dwarf. With a rich, unresolved pulsation spectrum
and the opportunity to learn some very interesting physics via 
asteroseismology, KPD~1930 was chosen for observation by the Whole Earth
Telescope (WET). KPD~1930 also has an infrared companion $\approx 0.5"$ away 
(\O stensen, Heber, Maxted 2005).

\section{Observations}
KPD~1930 was the target of the WET run Xcov~23.
Nearly 355 hours of data were collected at 17 observatories from 15 August to
9 September, 2003. The individual runs are provided in Table~\ref{tab01}.
Overall, these data have an observational duty-cycle of 36\% 
which is less than typical WET campaigns.
Because of the crowded field, most of the data were obtained with CCD
photometers, but some data were obtained with photoelectric (PMT) photometers.
The photoelectric data were reduced in the usual manner as described
by Kleinman, Nather, \& Phillips (1996). 
The standard procedures of CCD image reduction,
including bias subtraction, dark correction and flat field
correction, were followed using IRAF packages.
Differential intensities were determined via aperture photometry with
the aperture optimized for each individual run with varying numbers of
comparison stars depending on the field of view. 

We will also examine a small multisite campaign that obtained data during
July, 2002. In total, almost 45 hours of data were obtained 
from McDonald (the 2.1~m Otto Struve Telescope),
San Pedro-Martir (1.5~m) and Suhora (0.6~m) observatories.
Specifics of these runs are in Table~\ref{tab02}.

\begin{table*}
\centering
\caption{WET observations of KPD~1930+2752 during XCov23 in 2003 \label{tab01}}
\begin{tabular}{|lcrl|lcrl|} \hline
Run & Length & Date & Observatory & Run & Length & Date & Observatory\\
 & (Hrs) & UT & & & (Hrs) & UT &  \\ \hline
t081403 & 5.5 &  15 Aug. & Mt. Cuba 0.4m & loi2708 & 3.0 &  27 Aug. & Loiano 1.5m \\
phot081503 & 6.4 &  16  Aug.& Mt. Cuba 0.4m & gv30808 & 5.8 &  27 Aug. & OHP 1.9m \\
phot081703 & 0.7 &  18 Aug. & Mt. Cuba 0.4m & a0688 & 2.0 &  28 Aug. & McDonald 2.1m \\
hunaug18 & 3.1 &  18 Aug. & Piszkesteto 1.0m & mdr245 & 0.9 &  28 Aug. & KPNO 2.1m \\
NOT\_Aug19 & 7.8 &  19 Aug. & NOT 2.6m & haw28aug & 3.0 &  28 Aug. & Hawaii 0.6m \\
phot081903 & 6.5 &  19 Aug. & Mt. Cuba 0.4m & lulin28aug & 7.5 &  28 Aug. & Lulin 1.0m \\
hunaug19 & 4.3 &  19 Aug. & Piszkesteto 1.0m & turkaug28 & 2.5 &  28 Aug. & Turkey 1.5m \\
NOT\_Aug20 & 8.6 &  20 Aug. & NOT 2.6m & jr0828 & 3.7 &  28 Aug. & Moletai 1.65m \\
lna20aug & 4.4 &  20 Aug. & LNA 0.6m & retha-0031 & 3.4 &  28 Aug. & SAAO 1.9m \\
phot082003 & 6.9 &  20 Aug. & Mt. Cuba 0.4m & phot082903 & 4.9 &  29 Aug. & Mt. Cuba \\
hunaug20 & 4.6 &  20 Aug. & Piszkesteto 1.0m & a0690 & 0.2 &  29 Aug. & McDonald 2.1m \\
hunaug21 & 4.8 &  21 Aug. & Piszkesteto 1.0m & mdr246 & 3.8 &  29 Aug. & KPNO 2.1m \\
NOT\_Aug21 & 8.8 &  21 Aug. & NOT 2.6m & haw29aug & 3.0 &  29 Aug. & Hawaii 0.6m \\
lna21aug & 4.5 &  21 Aug. & LNA 0.6m & turkaug29 & 5.5 &  29 Aug. & Turkey 1.5m \\
lulin21aug & 7.0 &  21 Aug. & Lulin 1.0m & retha-0041 & 3.9 &  29 Aug. & SAAO 1.9m \\
NOT\_Aug22 & 8.7 &  22 Aug. & NOT 2.6m & gv30809 & 5.7 &  29 Aug. & OHP 1.9m \\
lna22aug & 1.3 &  22 Aug. & LNA 0.6m & mdr247 & 6.8 &  30 Aug. & KPNO 2.1m \\
haw22aug & 1.0 &  22 Aug. & Hawaii 0.6m & haw30aug & 2.4 &  30 Aug. & Hawaii 0.6m \\
lulin22aug & 2.4 &  22 Aug. & Lulin 1.0m & turkaug30 & 4.3 &  30 Aug. & Turkey 1.5m \\
suh-114 & 3.9 &  22 Aug. & Suhora 0.6m & retha-0051 & 3.9 &  30 Aug. & SAAO 1.9m \\
gv30801 & 2.5 &  22 Aug. & OHP 1.9m & loi0830 & 0.1 &  30 Aug. & Loiano 1.5m \\
NOT\_Aug23 & 6.5 &  23 Aug. & NOT 2.6m & mdr248 & 6.5 &  31 Aug. & KPNO 2.1m \\
lna23aug & 3.1 &  23 Aug. & LNA 0.6m & haw31aug & 2.9 &  31 Aug. & Hawaii 0.6m \\
haw23aug & 3.6 &  23 Aug. & Hawaii 0.6m & turkaug31 & 5.5 &  31 Aug. & Turkey 1.5m \\
lulin23aug & 6.8 &  23 Aug. & Lulin 1.0m & se0103q1 & 2.6 &  01 Sep. & SSO 1.0m \\
gv30803 & 1.1 &  23 Aug. & OHP 1.9m & turksep01 & 5.0 &  01 Sep. & Turkey 1.5m \\
lna24aug & 4.8 &  24 Aug. & LNA 0.6m & mdr249 & 3.6 &  02 Sep. & KPNO 2.1m \\
phot082403 & 6.9 &  24 Aug. & Mt. Cuba 0.4m & se0203q1 & 4.6 &  02 Sep. & SSO 1.0m \\
haw24aug & 0.5 &  24 Aug. & Hawaii 0.6m & turksep02 & 4.9 &  02 Sep. & Turkey 1.5m \\
gv30805 & 3.5 &  24 Aug. & OHP 1.9m & haw03sep & 2.3 &  03 Sep. & Hawaii 0.6m \\
phot082503 & 3.8 &  25 Aug. & Mt. Cuba 0.4m & se0303q1 & 4.5 &  03 Sep. & SSO 1.0m \\
haw25aug & 0.7 &  25 Aug. & Hawaii 0.6m & wise03sep & 6.0 &  03 Sep. & Wise 1.0m \\
lulin25aug & 0.2 &  25 Aug. & Lulin 1.0m & haw04sep & 2.4 &  04 Sep. & Hawaii 0.6m \\
suh-116 & 3.1 &  25 Aug. & Suhora 0.6m & se0403q1 & 4.6 &  04 Sep. & SSO 1.0m \\
gv30806 & 4.5 &  25 Aug. & OHP 1.9m & wise04sep & 3.8 &  04 Sep. & Wise 1.0m \\
phot082603 & 6.6 &  26 Aug. & Mt. Cuba 0.4m & hunsep04 & 1.7 &  04 Sep. & Piszkesteto 1.0m \\
a0684 & 0.5 &  26 Aug. & McDonald 2.1m & haw05sep & 2.5 &  05 Sep. & Hawaii 0.6m \\
a0685 & 0.3 &  26 Aug. & McDonald 2.1m & wise05sep & 7.0 &  05 Sep. & Wise 1.0m \\
lulin26aug & 7.0 &  26 Aug. & Lulin 1.0m & haw06sep & 2.2 &  06 Sep. & Hawaii 0.6m \\
retha-0020 & 0.7 &  26 Aug. & SAAO 1.9m & wise06sep & 7.0 &  06 Sep. & Wise 1.0m \\
gv30807 & 4.9 &  26 Aug. & OHP 1.9m & hunsep06 & 4.8 &  06 Sep. & Piszkesteto 1.0m \\
lna27aug & 3.5 &  27 Aug. & LNA 0.6m & haw07sep & 1.2 &  07 Sep. & Hawaii 0.6m \\
lulin27aug & 7.4 &  27 Aug. & Lulin 1.0m & hunsep07 & 4.6 &  07 Sep. & Piszkesteto 1.0m \\
retha-0021 & 3.7 &  27 Aug. & SAAO 1.9m & hunsep09 & 0.7 &  09 Sep. & Piszkesteto 1.0m \\ \hline
\end{tabular}
\end{table*}

\begin{table*}
\centering
\caption{Observations of KPD~1930+2752 during 2002 \label{tab02}}
\begin{tabular}{|lcrl|lcrl|} \hline
Run & Length & Date & Observatory & Run & Length & Date & Observatory\\
 & (Hrs) & UT & & & (Hrs) & UT &  \\ \hline
20710 & 3.5 &  10 July & Suhora 0.6m & A0302 & 4.0 &  14 July & McDonald 2.1m\\
20711 & 1.8 &  11 July & Suhora 0.6m & A0304 & 3.9 & 15 July & Suhora 0.6m\\
A0296 & 2.0 &  11 July & McDonald 2.1m & 20717 & 5.9 &  15 July & McDonald 2.1m\\
20712 & 5.3 &  12 July & Suhora 0.6m & sp1 & 2.4 &  17 July & Suhora 0.6m \\
A0300 & 0.3 &  13 July & McDonald 2.1m & 20720 & 2.2 & 17 July & S.P. Martir 1.5m\\
A0301 & 4.5 &  13 July & McDonald 2.1m & sp2 & 2.4 & 20 July & Suhora 0.6m \\
20714 & 1.4 &  14 July & Suhora 0.6m & 20715 & 4.2 & 20 July & S.P. Martir 1.5m\\
\hline
\end{tabular}
\end{table*}

As sdB stars are
substantially hotter than typical field stars, differential light
curves are not flat due to atmospheric reddening. A low-order
polynomial was fit to remove nightly trends from the data.
Finally, the lightcurves were normalized by their average flux and
centered around zero so the reported differential intensities are
$\Delta I = \left( I /\langle I\rangle\right) -1 $. Amplitudes are
given as milli-modulation amplitudes (mma), with 10~mma
corresponding to 1.0\% or 9.2~millimagnitudes.

\begin{figure*}
 \centerline{
\psfig{figure=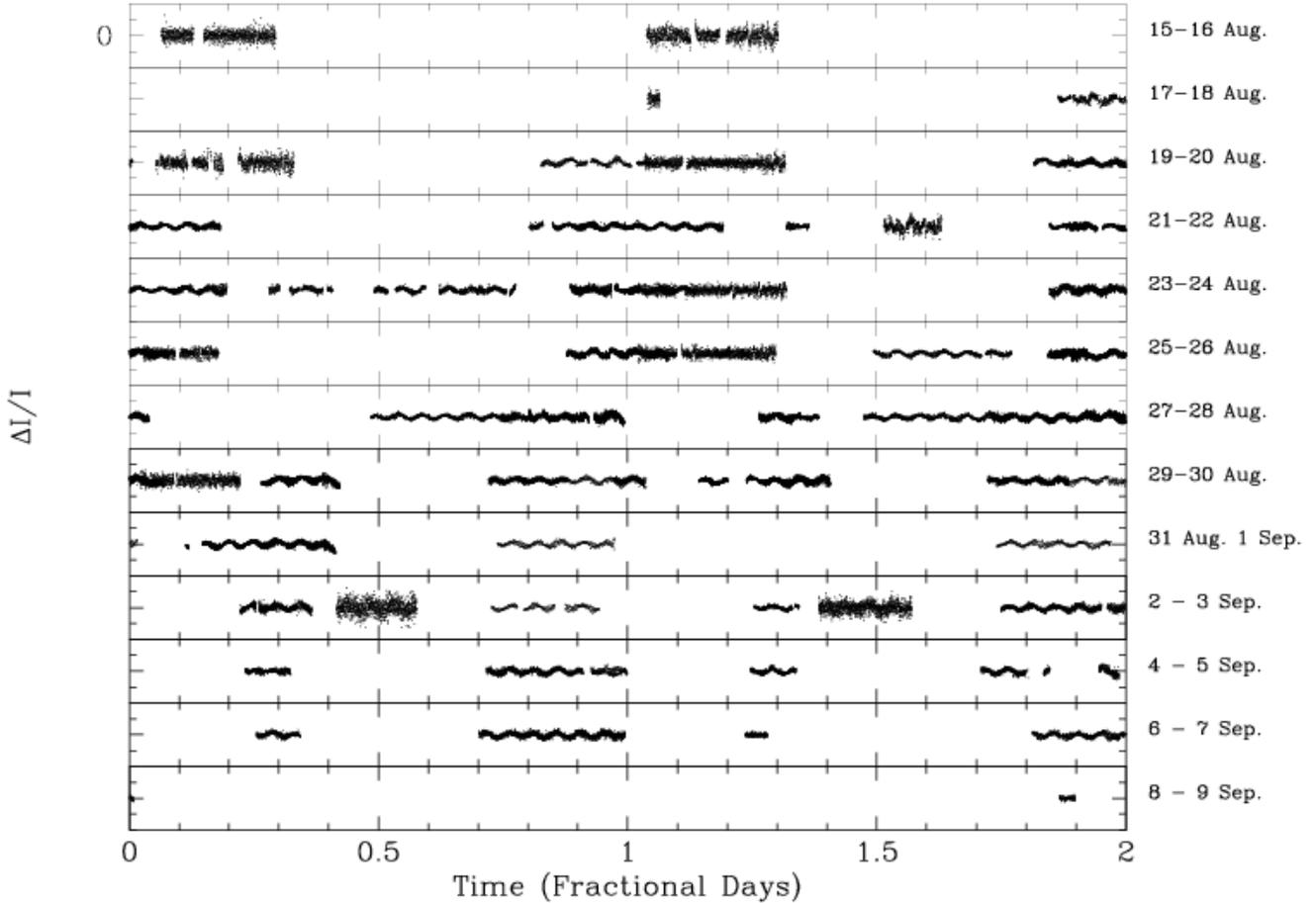,width=\textwidth}}
 \caption{Light curves showing all data obtained during Xcov~23.
Each panel is 2 days with the dates given on the right. \label{fig01}}
\end{figure*}


\begin{table}
\centering
\caption{Various groups of data used in our pulsation analysis. Columns 3 and
4 indicate the temporal resolution (in $\mu$Hz) and $4\sigma$ detection limit
(in mma).
\label{tab03}}
\begin{tabular}{lccc}\hline
Group & Inclusive dates & Res. & Limit \\ \hline
I & 15 Aug. - 9 Sep. & 0.45 & 0.44\\
II & 18 Aug. - 7 Sep. & 0.60 & 0.31 \\
III & 19 - 23 Aug. & 2.69 & 0.44\\
IV & 27 - 31 Aug. & 2.58 & 0.52 \\
V & 3 - 6 Sept. & 3.75 & 0.77 \\
VI & 2002, 10 - 21 July & 1.87 & 0.59 \\ \hline
\end{tabular}
\end{table}

\section{Analysis}
\subsection{Orbital Parameters}
\begin{figure*}
 \centerline{
\psfig{figure=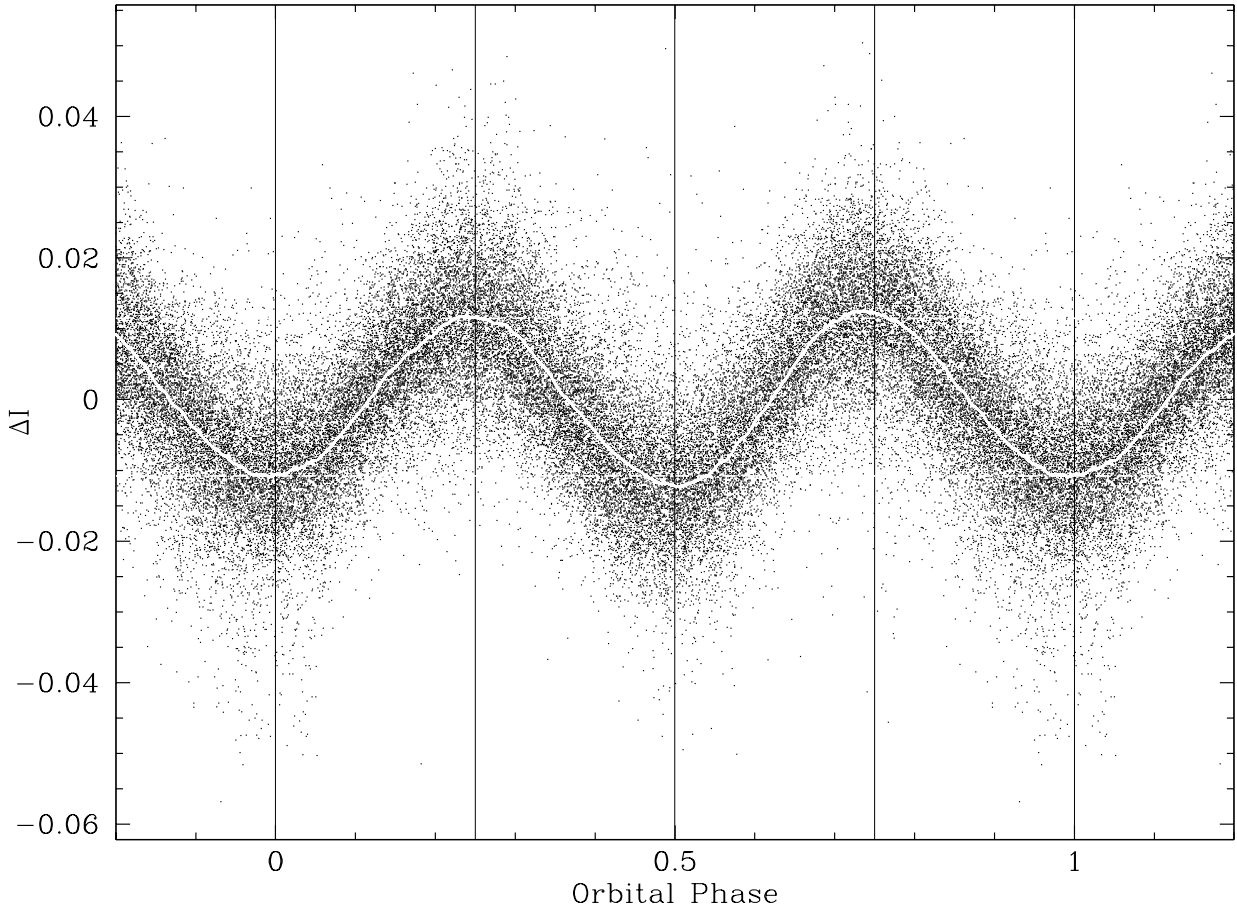,angle=-90,width=\textwidth}}
  \caption{XCov~23 data folded over the binary period. The solid line
is a result of prewhitening the data and smoothing over 335 points (60
seconds in time) and the dashed 
horizontal lines indicate one maximum and minimum.
For clarity, slightly more than one orbit is shown.
\label{figOrb} }
\end{figure*}

The largest variation in the lightcurve is caused by an ellipsoidal variation
of the sdB star. In removing this variation to examine the pulsations, we
can deduce some of the orbital properties. We defer attempting a complete
binary solution to a work in progress (Pablo et al. private communication),
but rather provide some basic information that are 
obvious from the data. 
Non-linear least-squares (NLLS) fitting
to the data (2002 and 2004) provides a frequency of ellipsoidal variation of
$243.36987\pm 0.00007\,\mu$Hz. The binary period is twice that at
$8217.994\pm 0.002\,s$ or $0.09511567\pm 0.00000003\,d$.  Our value is within
that found by B00 but outside the errors of the period determined by
Geier et al. (2007; hereafter G07). 
G07 had 2900 spectroscopic data points unevenly 
scattered throughout four years while we have 36\% coverage during 26 days
in 2003 and $\sim 45$ hours spanning 7 days in 2002. As the epochs of our
data overlap, period change can be ruled out and it is most likely that
one (or both) of us are underestimating our errors.
Figure~\ref{figOrb} shows modified
XCov~23 data folded over the binary period. The pulsations make the lightcurve
very broad and it even appears that a pulsation frequency is an integer
multiple of the binary period. However that is not the case, but rather there
are many pulsation frequencies between 33 and 34 times the orbital frequency
(see \S 3.2). 
To transform the broad pulsation-included lightcurve into a narrower,
prewhitened form, we used our best data (Group II), prewhitened by all
61 pulsation frequencies. Then we phase-folded the data over the orbital 
period, did a 60-point ($\approx 11.5$~s) smoothing, and fitted an additional
three frequencies. We then prewhitened these three frequencies from the 
original, non-phase folded data, phase folded again, and smoothed by
60 ($\approx 11.5$~s), 168 ($\approx 30$~s), and 335 ($\approx 60$~s)
points. The differently smoothed data did not affect the maxima and minima
of the orbital variations, and so we used the 335-point smoothed data,
shown as a line in Fig.~\ref{figOrb}, to examine the ellipsoidal variations.

Using radial velocity data, G07 estimate the orbital inclination
as $i\sim 80^o$. At that inclination, the white dwarf companion should
eclipse the sdB star. An eclipse is shown for $i= 80^o$ in Fig.~8 of G07 
and using the orbital parameters of G07, we produced a simulated lightcurve
with Binary Maker 3\footnote{See www.binarymaker.com.}. 
Our simulation was not intended to
fit the actual data, but rather to illustrate the eclipse shape.
From the simulation, the eclipse duration would be $\approx 164$~s, or nearly
three of our binned data points (at a binning of 335 points) with a depth
of 0.2\%.
While the observed minima are uneven, the shape and depth do not match
that of an eclipse. We can therefore rule out inclinations for which an
eclipse should occur; namely $i\geq 78.5^o$, based on G07. The uneven
minima indicates what would be expected based on the
tidal distortion (gravity darkening). 
The gravity of the white dwarf makes the sdB star very
slightly egg-shaped. When the marginally pointier end is facing us, the 
line-of-sight angle does not go as deep at one optical
depth as it does when the blunter end is facing us. This produces a fainter
minimum when the sdB is furthest from us and we view the slightly 
pointier end. A vastly exaggerated schematic of this is shown in 
Fig.~\ref{figOrb3} (a reconstructed image based on Fig. 2 of Veen et al. 2002). 
We fitted regions spanning 0.2 in phase with Gaussians to determine each
maximum and minimum.
The minimum at $\phi = 0.5$ is $0.18\pm 0.03$\% fainter than the other one 
and using
detailed models, this information, along with the amplitude of the ellipsoidal
variation should be sufficient to constrain the shape and limb darkening of
KPD~1930. The uneven minima allows us to
define closest approach of the sdB star,
and like Maxted et al. (2000), we define that as the zero-point of the
orbital phase.

\begin{figure}
 \centerline{
\psfig{figure=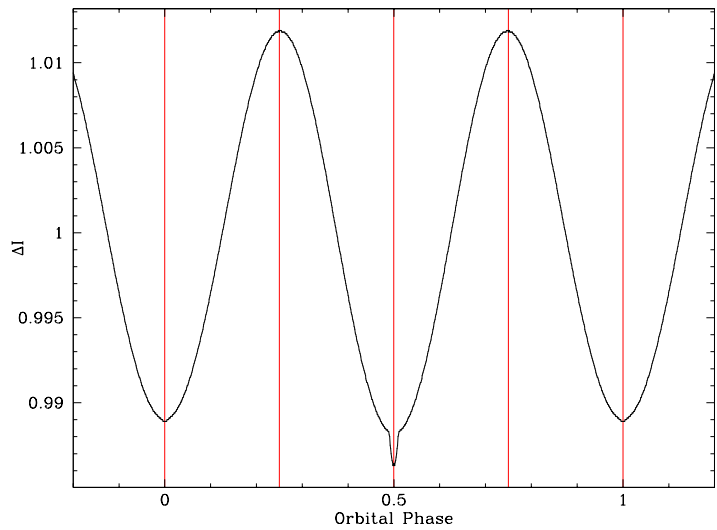,angle=-90,width=4.0in}}
  \caption{Simulated data showing the eclipse shape for $i=80^o$.
\label{figOrb2} }
\end{figure}

\begin{figure}
 \centerline{
\psfig{figure=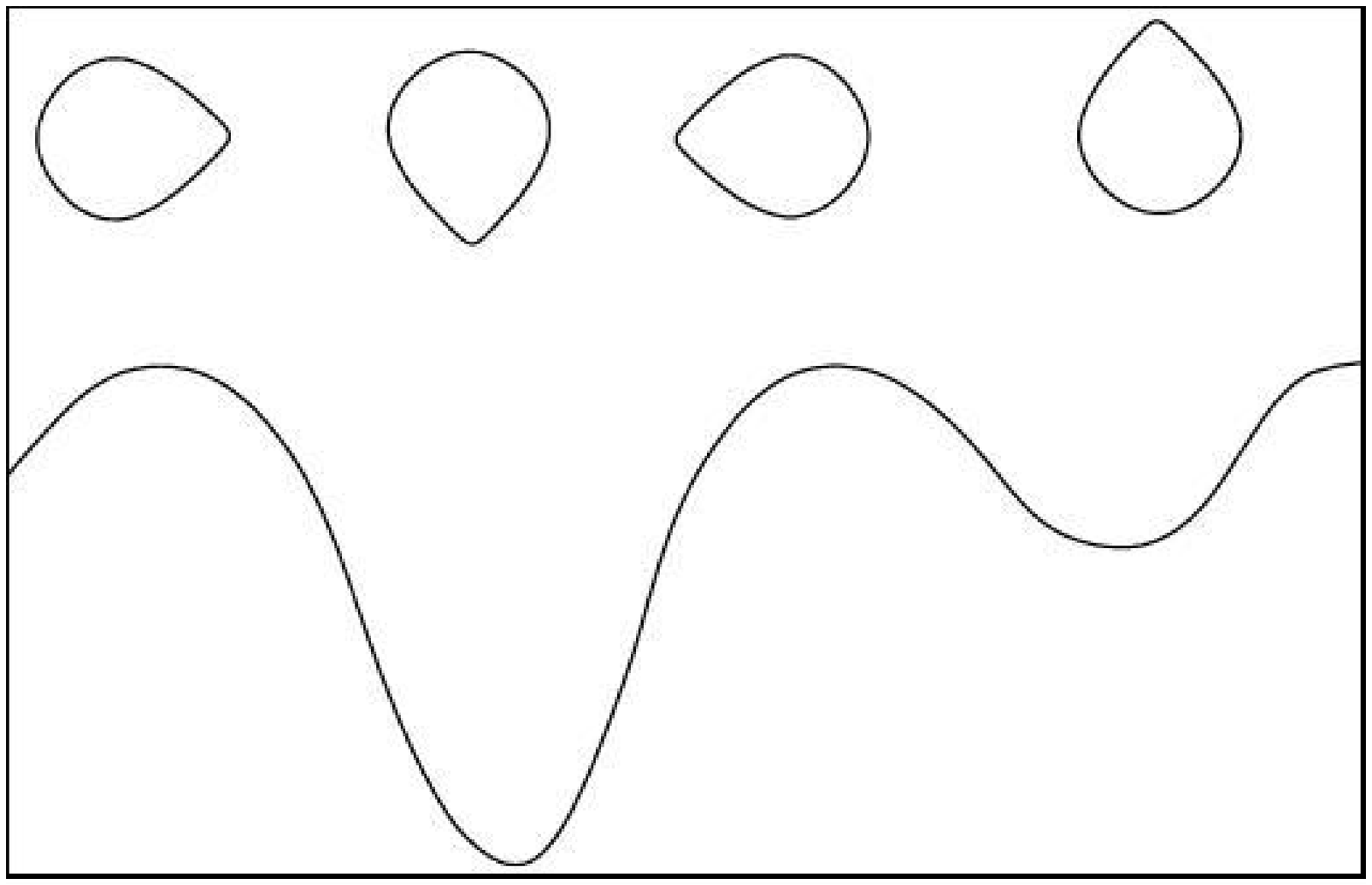,width=4.0in}}
\caption{Exaggerated simulated data showing how limb darkening, combined
with ellipsoidal variation causes uneven minima.
\label{figOrb3} }
\end{figure}

The lightcurve maxima are also not quite even. Using the same Gaussian
fitting technique, the maximum at $\phi_{orb}=0.75$, which
is when the sdB component is traveling towards us, is 
$0.06\pm 0.05$\% brighter 
then at $1.25$, when the star is traveling away from us.
Doppler boosting and Doppler shifting affects the flux by a factor of
$(1-v(t)/c)$ (G07; Maxted et al. 2000) which, using K1 from G07, 
should result in a Doppler
brightening semi-amplitude of 0.11\%. This is at the $1\sigma$ 
upper-limit of our measurement. Instead of using the measured
velocity to deduce flux differences in maxima, two of us (SDK and HP)
attempted to constrain the velocity from the lightcurve itself. Using
low-pass filtering of the light curve, we obtained approximate values
for the flux increase and converted that to velocity. The estimate
of the  maximum radial velocity is $529.76 \pm 16\,km/s$ 
which is the same order as the radial
velocity measurements of G07. The error estimate is the error of the fit 
to the light curve and does not include the (large) systematic error 
of the spectrum estimate. More on this measurement and its effects on the 
binary  will be discussed in a future paper (Pablo et al. in preparation).
So Doppler effects could be responsible 
for the difference in maxima. Interestingly, such special-relativistic 
effects have been detected in another sdB binary at the expected level
(Bloemen et al., in press).

\subsection{Pulsation analysis}
\begin{figure*}
 \centerline{
\psfig{figure=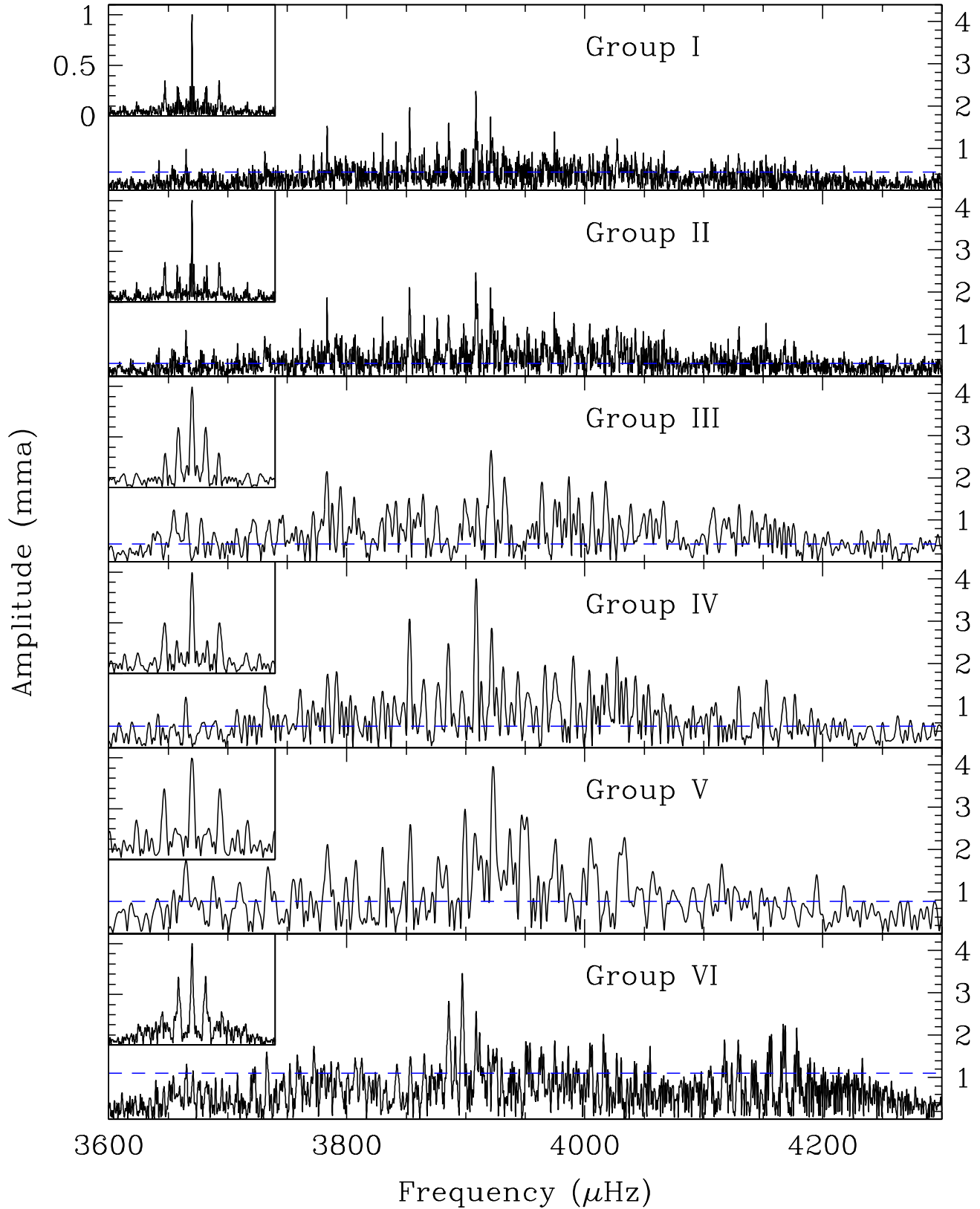,width=\textwidth}}
  \caption{Plot of the temporal spectra and window functions (inset) for 
the groupings of data in Table~\ref{tab03} for the frequency range of
3600 to 4250~$\mu$Hz. Dashed lines are the $4\sigma$ detection limit.
\label{fig02}}
\end{figure*}

\begin{figure*}
 \centerline{
\psfig{figure=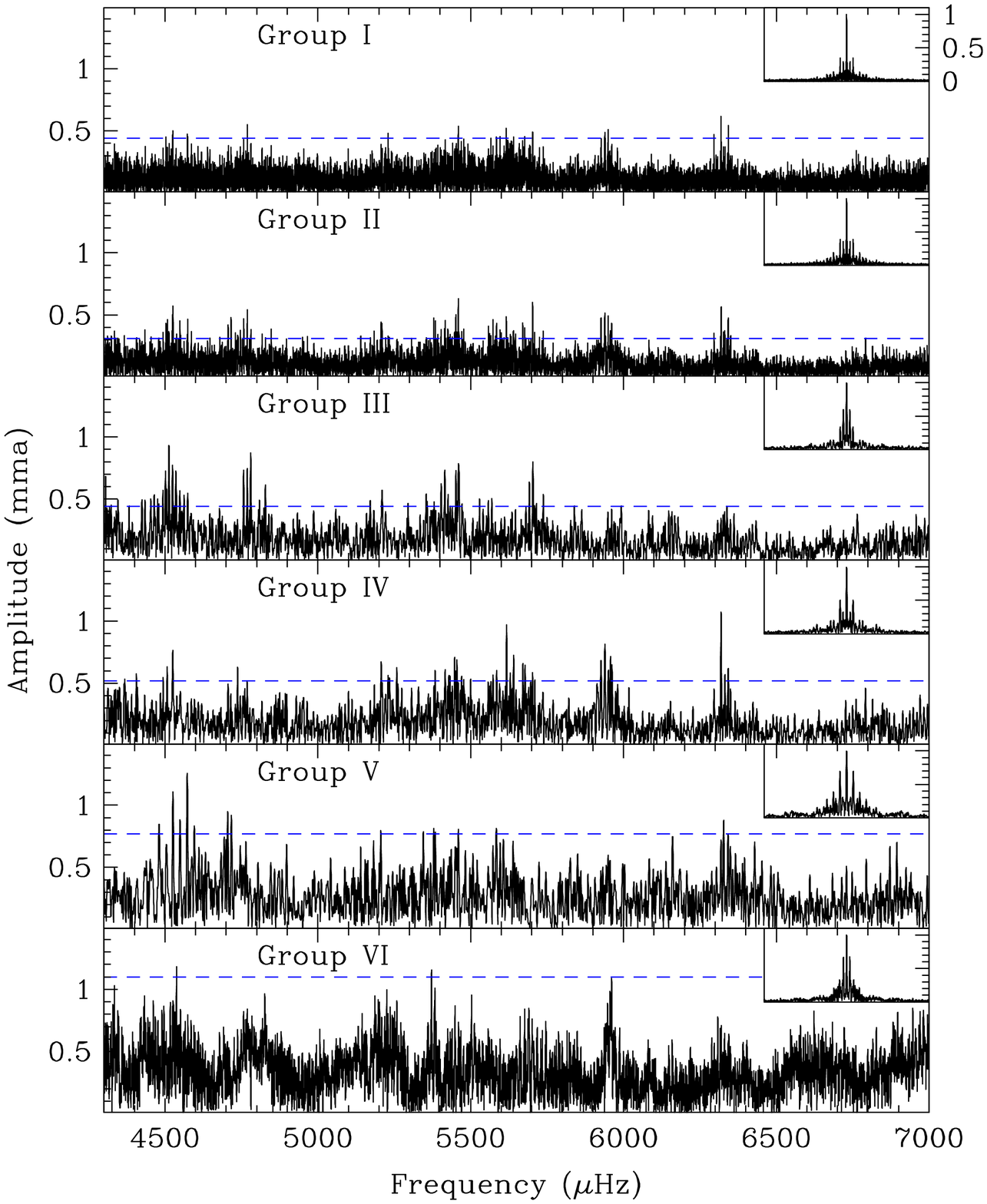,width=\textwidth}}
  \caption{Same as Fig.~\ref{fig02} for the frequency range of
4300 to 7000~$\mu$Hz. \label{fig03}}
\end{figure*}

In a standard manner for long time-series data (i.e. Kilkenny et al. 1999; Reed 
et al. 2004b, 2007b), we analyzed the data in several different groupings.
We can use these groups to examine frequency and amplitude stability,
look for consistent frequencies, and amplitude variations.
The groups are provided in Table~\ref{tab03} and pertinent sections
of their temporal
spectra (Fourier transforms; FTs) are shown in Figs.~\ref{fig02} and 
\ref{fig03}. Table~\ref{tab03} also lists the temporal resolution (defined as
$1/t$ where $t$ is the run duration) and the $4\sigma$ detection limit
determined for the range from 1~000 -- 3~000 and 8~000 -- 10~000~$\mu$Hz.
 Group I data 
includes all of the XCov~23 data.
Group II excludes noisy data, which is 
defined as a $4\sigma$ limit above 2~mma, and has 
been trimmed so that for any overlapping
data, only the best quality data were kept.
Groups III, IV, and V contain data obtained over 4 or 5 days of 
relatively good coverage during three different weeks of the 2003 campaign.
Group VI is the 2002 small multisite campaign. Figure insets show the
window functions, which are a single sine wave temporally sampled as the actual
data. The central peak is the input frequency while other peaks are aliases
which can complicate the data. Groups I and II are relatively well-sampled,
with alias peaks less then 40\% of the input amplitude whereas the remaining
groups have obvious aliases that will contribute to the overall noise of
the data.

A glance at the figures provides two simple observations: 
i) KPD~1930 is multiperiodic, pulsating in several
tens of modes, confirming what was found in the discovery data (B00).
ii) The pulsation frequencies are short-lived. This is evident in that the
highest-amplitude peaks change between the shortest data sets (Groups
III through VI) and the groups with the longest data sets (I and II) 
have the lowest amplitudes.

\begin{figure*}
 \centerline{
\psfig{figure=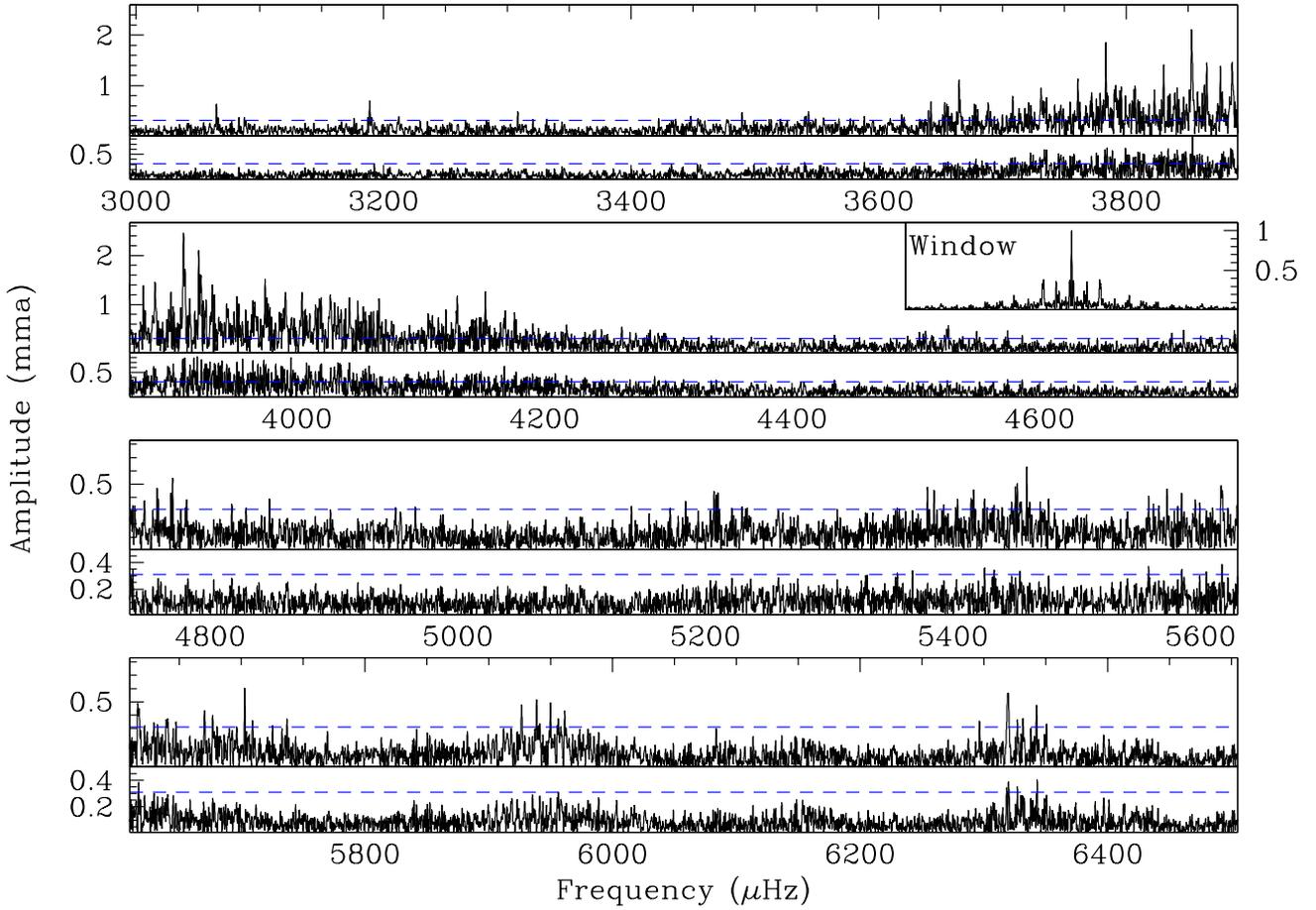,angle=-90,width=\textwidth}}
  \caption{A detailed FT of Group II data showing the residuals after
prewhitening by 61 frequencies. Each panel shows the same frequency resolution,
but the amplitudes in the top two panels differ from the bottom two. The
dashed (blue) line is the $4\sigma$ detection limit.
\label{fig05} }
\end{figure*}

As peak amplitudes in 
FTs show a mixture of the \emph{median} amplitude and effects of phase
stability, prewhitening of the combined data sets could not be
expected to accurately remove variations in amplitudes and/or
phases (see  Koen 2009 and Reed et al. 2007a). 
However, because of the frequency density shorter duration data
sets would likely have unresolved frequencies. 
This does not mean that least-squares fitting and prewhitening
are not of use, just that they need to be used with caution. 
We did the usual method of simultaneously fitting and prewhitening 
the data using
non-linear least-squares (NLLS) software independently for all the groups
in Table~\ref{tab03} until we could not discern peaks (as in 
Fig.~\ref{fig05}). However, because of the rich pulsation spectrum we added
an additional step. When searching the FT for pulsation peaks, we also plotted
a window function made from the 
prewhitening information gathered for all fitted frequencies. In this manner,
we could view cumulative windowing effects and better judge the
impact of prewhitening on the FT. (A prewhitening sequence of Group~IV's data
is provided as Fig.~1 in on-line supporting materials.)

Frequencies were deemed intrinsic to the star if they were
detected in at least two Groups above the $4\sigma$ detection limits.
Table~\ref{tab05} provides a list of detected 
pulsation frequencies.
The first column lists a designation,
the second gives the frequency from the highest temporal-resolution Group and
the NLLS error in parentheses. Subsequent columns list the fitted amplitude
from each group (NLLS error in parentheses) along with any pertinent notes:
NF indicates
that the frequency was not NLLS fitted; and frequencies fitted, but
off by a daily aliases are noted as $+$ or $-$ for one daily alias away
or $++$ or $--$ for two daily aliases away. The last column notes other
frequencies that are within $1\,\mu$Hz of the daily alias as this could make
prewhitening difficult, depending on the window function. Table~\ref{tab06}
lists other frequencies that we \emph{suspect} are intrinsic to the star,
but did not meet our requirements. Figure~\ref{fig05} is an expanded FT
for Group~II's data. Each two-panel section shows the
original FT and the residuals after prewhitening. The dashed (blue) line 
is the $4\sigma$ detection limit. There are regions where the residuals
remain higher then the $4\sigma$ detection limit because either NLLS fitting
and prewhitening did not effectively remove all of the amplitude (most
likely caused by amplitude and/or phase variations or another, 
unresolved frequency) or
because remaining peaks were not observable in the original FT.


\begin{table*}
\caption{Frequencies and amplitudes detected in various groupings of data. The temporal resolution of
1/(run length) and the $4\sigma$ detection limit are provided in the first two 
rows.
Frequencies are in $\mu$Hz and amplitudes in mma with NLLS fitting errors 
on the last digits in parentheses. The last column indicates other frequencies
that are within 1~$\mu$Hz of the daily alias. Notes: + indicates
that the frequency fit using NLLS was 1 daily alias ($11.5\mu$Hz) larger 
than that listed. Likewise, ++, -,
and $--$ indicate frequencies that are plus 2 daily, minus 1 daily, and minus 2 daily aliases from that listed,
respectively. $\approx$ indicates frequencies were slightly more than $1\sigma$ 
away from that
shown, $<4\sigma$ indicates frequencies that had power in the FT which 
was below  the $4\sigma$
detection limit, NF indicates frequencies above the $4\sigma$ limit that could 
not be fit using
our NLLS program.
 $\dagger$ indicates unresolved frequencies. 
Amplitudes of unresolved frequencies do not reflect the intrinsic amplitudes
of each frequency. $\diamond$ indicates frequencies that match B00 while 
$\star$ indicates those that are a daily alias away. \label{tab05}}
{\footnotesize
\begin{tabular}{|l|c|c|c|c|c|c|c|c|c|} \\ \hline
 & Group & I         &  II           & III           & IV       & V    & VI \\ \hline
 & $4\sigma$ limit & 0.44 & 0.31 & 0.44 & 0.52 & 0.77 & 0.59 \\ \hline
 & Resolution & 0.45 & 0.60 & 2.69 & 2.58 & 3.75 & 1.87 \\ \hline
ID & Frequency      & \multicolumn{6}{c}{Amplitudes} & Alias  \\ \hline
$f1$ & 3065.085 (49) & 0.70 (8) & 0.62 (8) & 0.56 (13) & 0.75 (13) & .. & 0.79 (
14) \\ \hline
$f2^{\star}$ & 3188.864 (44) & 0.68 (8) & 0.69 (8) & 0.62 (13) & 0.88 (13) & $<4\sigma$
 & 0.61 (14) \\ \hline
$f3$ & 3308.382 (66) & 0.44 (8) & 0.45 (8) & NF & $<4\sigma$ & $<4\sigma$+ & 0.63 
(14)+ \\ \hline
$f4$ & 3422.386 (92) & .. & 0.32 (8) & .. & .. & .. & 0.56 (14)+ \\ \hline
$f5$ & 3489.608 (80) & 0.48 (8) & 0.38 (8) & .. & $<4\sigma$ & .. & .. \\ \hline
$f6$ & 3543.258 (60) & 0.52 (8) & 0.50 (8) & ..  & 0.75 (13) & .. & .. \\ \hline
$f7$ & 3653.074 (54) & 0.55 (8) & 0.59 (9) & 1.13 (14) & .. & .. & .. & f8 \\ \hline
$f8$ & 3664.968 (29) & $\approx$0.92 (9) & 1.10 (9) & .. & 1.09 (14) & 1.91 (24)
 & 1.25 (15) & f7\\ \hline
$f9$ & 3683.269 (63) & 0.41 (8) & .. & $\approx$0.95 (14) & .. & .. & 1.17 (16) \\ \hline
$f10^{\diamond}$ & 3731.065 (38) & 0.84 (9) & 0.79 (8) & 1.40 (14)+ & 1.08 (15) & 1.15 (24)
++ & 1.06 (16)- \\ \hline
$f11^{\diamond}$ & 3783.460 (17) & 1.51 (9) & 1.88 (9) & 2.35 (14) & 1.84 (15) & 1.52 (26)
 & 1.55 (19) \\ \hline
$f12$ & 3791.545 (41) & NF & 0.79 (9) & .. & NF & .. & 1.15 (19)+ \\ \hline
$f13^{\diamond}$ & 3804.693 (41) & .. & 0.79 (9) & .. & .. & .. & 1.67 (17) \\ \hline
$f14$ & 3822.646 (39) & 0.70 (9) & NF & 1.28 (14)+NF & 0.90 (15)++ & $\approx$1.96 (
26) & .. \\ \hline
$f15^{\star}$ & 3852.709 (19) & 1.64 (9) & 1.90 (9) & 1.63 (14) & 2.27 (15) & NF & ..
& f16 \\ \hline
$f16$ & 3862.952 (34) & .. & 1.06 (9) & NF & .. & .. & 1.57 (16) & f15 \\ \hline
$f17^{\star}$ & 3908.614 (17) & 2.26 (11) & 2.09 (9) & .. & 3.76 (15) & .. & 2.85 
(17)- & f18 \\ \hline
$f18$ & 3920.717 (27) & 1.07 (10) & 1.40 (10) & NF & NF & NF & 1.04 (16)$\dagger$
& f17 \\ \hline
$f19$ & 3921.901 (28) & 1.22 (9) & 1.29 (9) & 2.40$\dagger$ (14) & .. & .. & .. \\ \hline
$f20$ & 3924.426 (26) & 1.30 (9) & 1.24 (9) & .. & .. & 3.94 (25)$\dagger$ & .. \\ \hline
$f21$ & 3926.008 (25) & 1.11 (9) & .. & .. & .. & .. & 1.77 (16) \\ \hline
$f22^{\star}$ & 3958.804 (35) & $\sim$0.66 (9) & 0.94 (9) & .. & .. & .. & .. \\ \hline
$f23$ & 3964.686 (30) & 0.93 (9) & .. & 1.72 (15) & .. & .. & .. & f24 \\ \hline
$f24^{\diamond}$ & 3974.320 (19) & 1.60 (9) & 1.68 (9) & $\approx$1.29 (15) & 1.97 (15) 
& 3.16 (25) & 1.76 (16) & f23 \\ \hline
$f25^{\diamond}$ & 3990.848 (29) & 0.65 (9) & 1.12 (9) & $\approx$1.04 (16) & 1.99 (15) & $\approx$1.74 (25) & .. \\ \hline
$f26$ & 4004.192 (25) & .. & 1.31 (9) & .. & .. & 2.01 (25) & .. \\ \hline
$f27$ & 4018.289 (27) & 1.22 (9) & 1.19 (9) & 1.85 (14) & 1.51 (15) & .. & 
2.11 (16) & $s72$ \\ \hline
$f28$ & 4027.241 (27) & 1.04 (9) & .. & NF & 2.19 (16) & .. & .. &  \\ \hline
$f29$ & 4034.710 (31) & 0.90 (9) & .. & .. & 1.37 (16) & .. & .. \\ \hline
$f30$ & 4049.223 (29) & 0.86 (9) & 1.12 (9) & 1.23 (14) & 1.27 (15) & .. & .. \\ \hline
$f31^{\star}$ & 4066.443 (33) & 0.84 (9) & 0.98 (9) & 1.21 (14) & .. & NF & .. \\ \hline
\end{tabular}
}
\end{table*}
\begin{table*}
{\footnotesize
\begin{tabular}{|l|c|c|c|c|c|c|c|c|} \\ \hline
ID & Frequency & I         &  II           & III           & IV       & V    & VI \\ \hline
$f32^{\diamond}$ & 4120.459 (48) & 0.78 (9) & 0.65 (8) & .. & NF & $\approx$1.57 (24) & 1.41 (16)- \\ \hline
$f33$ & 4129.667 (32) & 0.69 (9) & 0.95 (9) & 1.34 (13) & .. & .. & 1.59 (16) \\ \hline
$f34^{\star}$ & 4152.155 (33) & 0.55 (9) & 0.92 (9) & $\approx$1.02 (13) & 1.42 (14) & .. & .. \\ \hline
$f35$ & 4168.331 (36) & 0.92 (9) & 0.85 (8) & .. & 1.19 (14) & .. & 1.81 (15)+ \\ 
\hline
$f36^{\diamond}$ & 4195.364 (51) & 0.65 (9) & 0.59 (8) & .. & 0.75 (14) & 1.03 (22) & 0.98 (14)
 \\ \hline
$f37^{\diamond}$ & 4262.566 (69) & .. & 0.44 (8) & .. & 0.57 (14) & .. & .. \\ \hline
$f38$ & 4297.849 (58) & 0.44 (8) & 0.43 (8)- & 0.76 (17) & 0.57 (13) & .. & .. 
& $s75$ \\ \hline
$f39^{\diamond}$ & 4453.34 (31) & .. & .. & 0.66 (13) & .. & .. & 0.62 (14)- \\ \hline
$f40$ & 4480.61 (57) & .. & .. & .. & .. & 0.78 (22) & 0.89 (14) \\ \hline
$f41$ & 4507.884 (62) & .. & 0.48 (8) & 0.50 (13)- & .. & NF & 0.57 (15)+ \\ \hline
$f42$ & 4524.708 (56) & 0.66 (8) & 0.53 (8) & 0.99 (13)- & 0.70 (13) & .. & 0.61
 (14) \\ \hline
$f43$ & 4549.152 (69) & .. & 0.44 (8) & 0.49 (13) & .. & .. & .. \\ \hline
$f44$ & 4572.412 (42) & 0.61 (8) & NF & 0.51 (13)- & .. & 1.2 (22) & .. \\ \hline
$f45$ & 4716.271 (67) & NF & 0.45 (8) & .. & 0.60 (13)++NF & 0.98 (22) & NF- \\ \hline
$f46$ & 4769.657 (67) & 0.53 (8) & 0.45 (8) & 0.58 (14)- & .. & .. & .. 
& $f47$ \\ \hline
$f47^{\diamond}$ & 4781.042 (94) & .. & 0.31 (8) & 0.67 (14) & .. & .. & .. & $f46$ \\ \hline
$f48$ & 4847.885 (84) & .. & 0.35 (8) & 0.55 (13)$--$ & .. & .. & 0.65 (14)- \\ \hline
$f49$ & 5184.009 (86) & .. & 0.34 (8) & .. & .. & .. & 0.64 (14) \\ \hline
$f50$ & 5207.451 (68) & NF & 0.44 (8) & NF & 0.67 (13)$\dagger$ & $<4\sigma$$\dagger$ & .. \\ \hline
$f51^{\star}$ & 5210.203 (78) & .. & 0.38 (8) & 0.50 (13)$\dagger$ & .. & .. & .. \\ \hline
$f52$ & 5230.003 (41) & 0.63 (8) & .. & .. & .. & NF++ & 0.51 (14)++ \\ \hline
$f53^{\star}$ & 5379.426 (79) & .. & 0.38 (8) & 0.57 (13)$--$ & .. & .. & .. \\ \hline
$f54$ & 5384.750 (83) & .. & 0.36 (8) & .. & 0.57 (13) & 0.71 (22) & 0.93 (14)+NF \\
 \hline
$f55^{\diamond}$ & 5416.641 (74) & .. & 0.40 (8) & 0.72 (13) & $\approx$0.65 (14)++ & .. & .. \\
 \hline
$f56$ & 5451.930 (57) & .. & 0.53 (8) & 0.72 (17) & 0.73 (14) & NF & $\approx$0.58 (
15) \\ \hline
$f57$ & 5459.529 (47) & 0.51 (8) & 0.64 (8) & $\approx$0.56 (17) & .. & $<4\sigma$ & 0.57 (15) \\ \hline
$f58^{\star}$ & 5529.22 (37) & .. & .. & 0.54 (13) & .. & .. & .. \\ \hline
$f59$ & 5572.909 (74) & .. & 0.40 (8) & $\approx$0.51 (13) & .. & 0.72 (22)+ & .. \\ \hline
$f60$ & 5616.391 (65) & 0.51 (8) & 0.46 (8) & .. & 0.94 (13) & NF & .. \\ \hline
$f61$ & 5670.540 (79) & .. & 0.38 (8) & .. & 0.64 (13) & .. & $\approx$0.61 (14) \\ \hline
$f62$ & 5702.901 (51) & NF & 0.59 (8) & 0.98 (13) & .. & .. & 0.79 (15) \\ \hline
$f63$ & 5709.319 (79) & .. & 0.38 (8) & .. & .. & .. & 0.56 (15)+ \\ \hline
$f64^{\diamond}$ & 5737.037 (83) & $<4\sigma$ & 0.35 (8) & 0.47 (13)+ & .. & .. & .. \\ \hline
$f65$ & 5938.582 (60) & 0.42 (9) & 0.54 (8) & .. & 0.74 (13) & .. & 0.51 (14)- \\ 
\hline
$f66^{\diamond}$ & 5950.308 (68) & 0.44 (9) & .. & $<4\sigma$+ & $\approx$0.61 (13) & .. & $\approx$0.74 (15) \\ \hline
$f67$ & 6319.148 (69) & 0.54 (9) & 0.44 (8) & .. & 1.03 (13) & 0.83 (22)+ & 1.40 (22
)+ \\ \hline
$f68$ & 6342.994 (62) & 0.52 (9) & 0.49 (8) & $<4\sigma$- & .. & .. & 0.75 (22) \\ \hline
ID & Frequency      & I         &  II           & III           & IV       & V 
  & VI \\ \hline
\end{tabular}
}
\end{table*}

\begin{table*}
\caption{Same as Table~\ref{tab06} for suspected frequencies. \label{tab06}}
{\footnotesize
\begin{tabular}{|l|c|c|c|c|c|c|c|c|c|} \\ \hline
 & Group & I         &  II           & III           & IV       & V    & VI
& Alias \\ \hline
$s69$ & 3448.361 (85) & .. & 0.35 (8) & .. & .. & .. & .. \\ \hline
$s70^{\diamond}$ & 3885.715 (28) & 1.03 (10) & .. & .. & .. & $\approx$NF & .. \\ \hline
$s71$ & 3995.235 (27) & 1.01 (9) & .. & .. & .. & NF & .. \\ \hline
$s72$ & 4030.437 (29) & 1.09 (9) & .. & .. & NF & NF & .. & $f27$ \\ \hline
$s73$ & 4232.67 (25) & .. & .. & 0.91 (14) & .. & .. & .. \\ \hline
$s74$ & 4246.23 (31) & .. & .. & 0.74 (14) & .. & .. & .. \\ \hline
$s75^{\star}$ & 4307.84 (29) & .. & .. & 0.87 (17) & .. & .. & .. & $f38$ \\ \hline
$s76$ & 4897.000 (78) & .. & 0.38 (8) & .. & .. & .. & .. \\ \hline
$s77$ & 4949.582 (94) & .. & 0.31 (8) & .. & .. & .. & .. \\ \hline
$s78$ & 4965.63 (10) & .. & 0.36 (8) & .. & .. & .. & .. \\ \hline
$s79^{\diamond}$ & 5140.42 (12) & .. & 0.25 (8) & .. & .. & .. & .. \\ \hline
$s80$ & 5769.48 (11) & .. & 0.28 (8) & .. & .. & .. & $\approx$0.55 (14) \\ \hline
$s81^{\star}$ & 6083.547 (83) & .. & 0.35 (8) & $<4\sigma$+ & .. & .. & .. \\ \hline

\end{tabular}
}
\end{table*}

\subsection{The frequency content}
In total, we detected 68 independent pulsation frequencies and 13 
suspected ones. Here we summarize some generalities which will be used
in subsequent sections. Only three frequencies are detected in all six data 
Groups while $f2$ nearly is. Six other frequencies are detected in five data
Groups and 13 are detected in four data Groups. 38 frequencies are within
a 650~$\mu$Hz region between 3650 and 4300~$\mu$Hz. Unlike most
sdBV stars, which contain several frequencies above 3~mma, KPD~1930 does
not have any in the integrated data (Groups I and II). Only $f17$ has an 
amplitude $>2\,mma$ in all detections, though it is only detected in four
Groups. 11 of the 68 frequencies have Group~I amplitudes $>1\,mma$
and these always have higher amplitudes in the shorter data sets (Groups
III through VI), if detected. These simple observations again lead to the
conclusion that pulsations of KPD~1930 are 
highly variable in amplitude and/or phase over the course of our observations.
Combine that with the density in the main region of pulsations and
accurate deciphering of the temporal spectrum will be difficult. Additionally,
the pulsations in uncrowded regions have low amplitudes which will
make them difficult to detect in short data sets.

\section{Discussion}
\subsection{Comparison with the discovery data}	
By combining their four nights of data, B00 obtained a resolution of
1.89~$\mu$Hz and their estimate of a mean noise level is 0.021\%,
giving their data a $4\sigma$ detection limit of 0.84~mma. By comparison, 
our XCov~23 data have $4.2\,\times$ better resolution and our detection
limit is about half.
Frequencies that match those of B00 are marked with
a $^{\diamond}$ in Tables ~\ref{tab05} and \ref{tab06} 
while those that are a daily alias away are marked with a $^{\star}$.
26 of our 81 frequencies are related to the 44 listed in B00. Of
the eight frequencies listed in B00 with amplitudes greater
than 2~mma, seven are detected in our data. For comparison, there are 21
frequencies which we have detected in at least five of our Groups or have
amplitudes $>1\,mma$ in at least two groups (one of which must be Group
I or II), and  of these, eight are related to frequencies detected in B00.
This also indicates a substantial amount of amplitude and/or phase
change since the B00 observations.

\subsection{Amplitude and phase stability}
KPD~1930 shows characteristics similar to the sdBV star PG~0048+091
(Reed et al. 2007b), which has pulsation properties normally associated
with stochastic oscillations. These properties include frequencies that are 
inconsistent between data sets and lower amplitudes in longer duration
data. Of the 69 frequencies in Table~\ref{tab05}, only $f10$, $f11$, and
$f24$ are detected in all six Groups of data, while $f2$ is detected in
five groups, with a peak just below $4\sigma$ in the sixth. Six more
frequencies are detected and fitted in five of the six groups. Unfortunately,
unlike PG~0048+091's well-spaced frequencies, most of those in KPD~1930 
are packed tightly between 3600 and 4200~$\mu$Hz. Outside of this main
region of power, the amplitudes are quite low, making detection difficult.

Despite these complications, we analyzed
data sets of varying length with the goal of reaching timescales
shorter than the timescale of amplitude and/or phase
variations. Such data would be free of the resultant complications,
allowing the amplitudes and phases to be more accurately
measured. These could then be examined over the duration of our observations 
for changes. For the shortest
possible time scale, we examined every individual run for which the
$4\sigma$ detection limit was better then 2.0~mma; totaling 32
runs. We also created 13 daily data 
sets combining low-noise runs that were contiguous, or nearly so. The
lengths of the daily data ranged from 7 to 27 hours, with a median value of
16.5 hours. Including the data for Groups III, IV, and V, we have data sets
sampling time scales near 0.25, 0.67, and 4 days. 

\begin{figure}
\psfig{figure=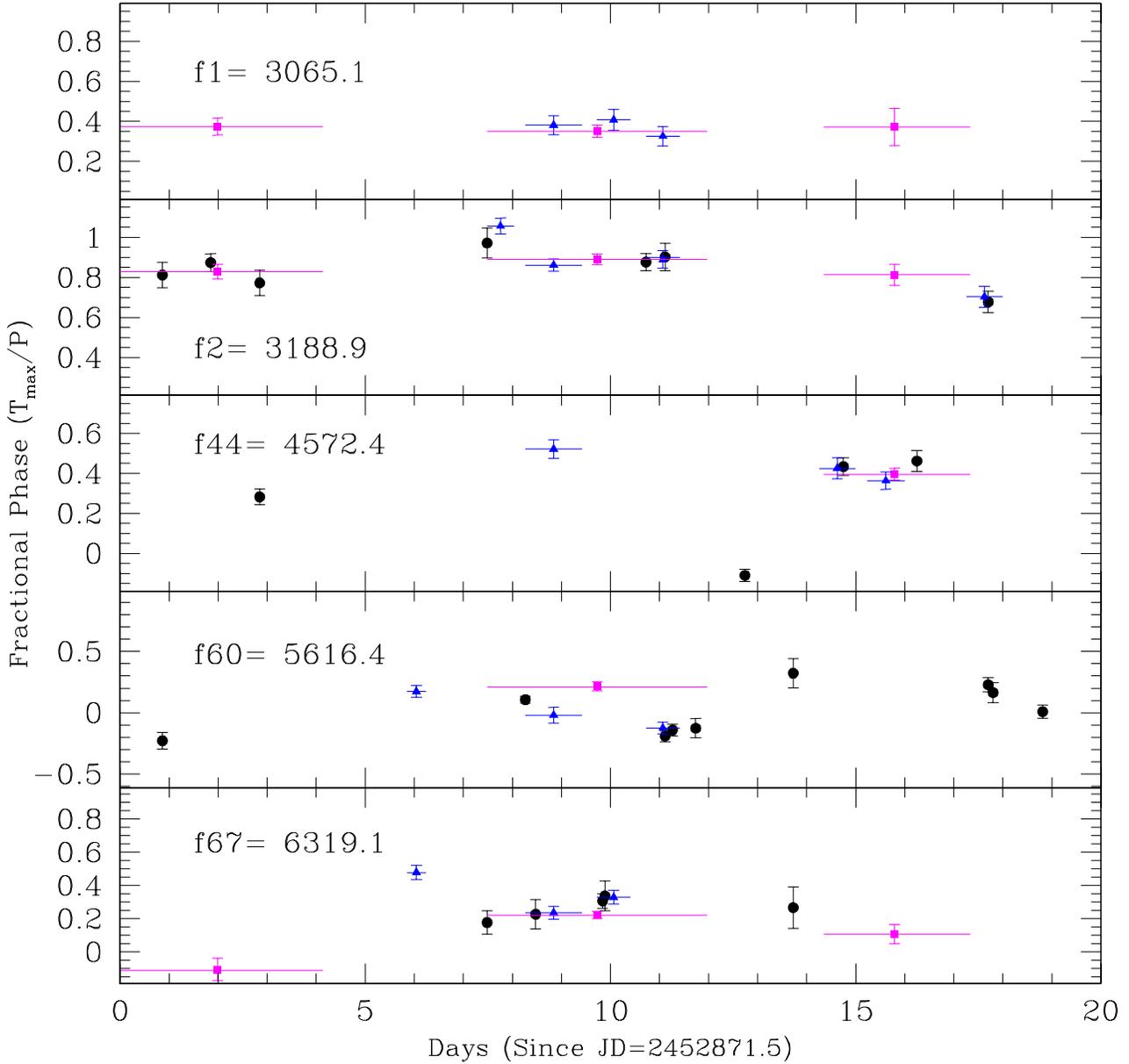,width=\textwidth}
\caption{Phases of frequencies resolvable from short data sets.
Circles (black) indicate individual runs, triangles (blue) are from daily
runs, and squares (magenta) are for Groups III -
V. Horizontal lines indicate the time span of the data used in the phase 
determination for the daily and group sets. \label{fig07} }
\end{figure}
\begin{figure}
\psfig{figure=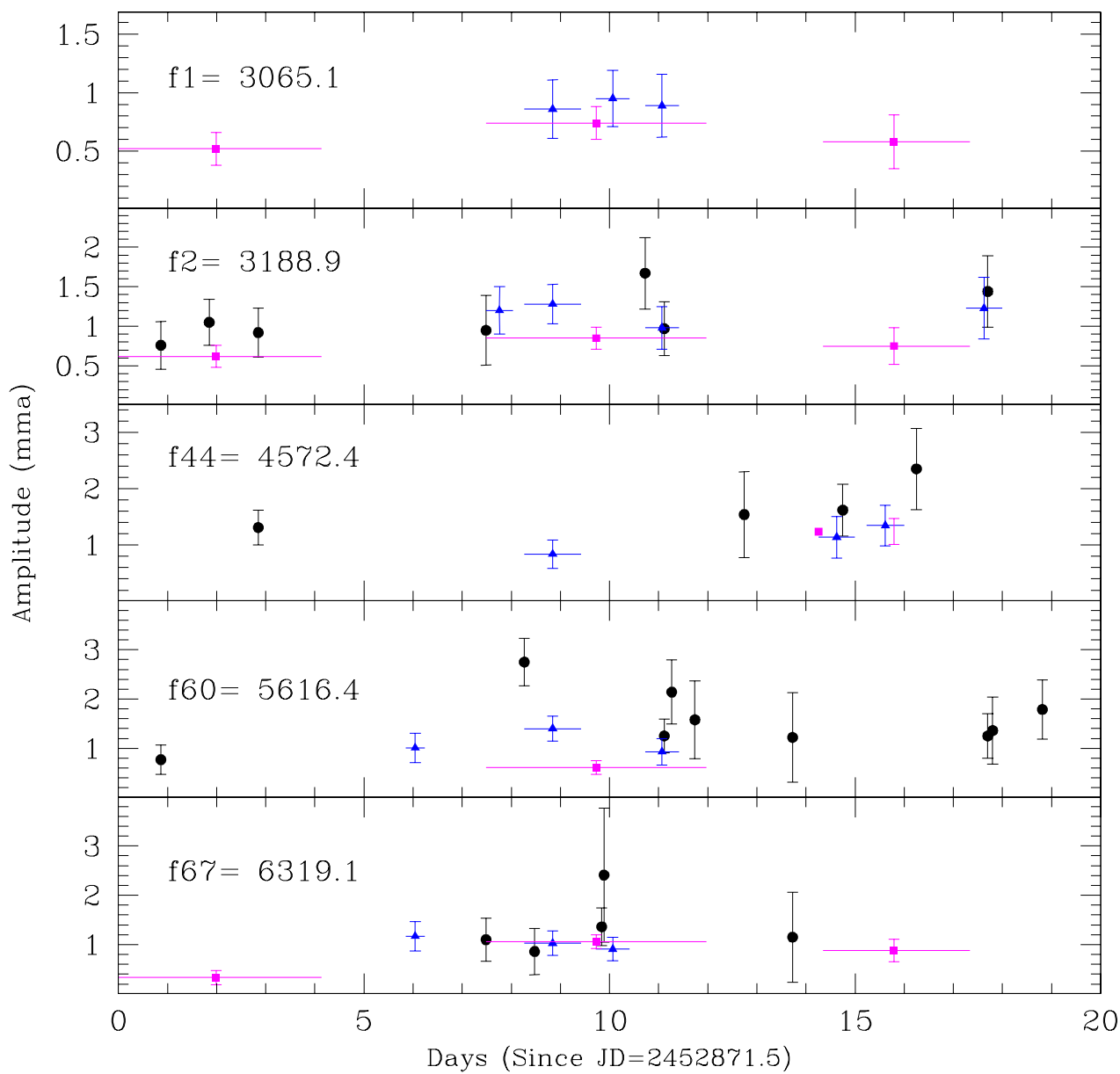,width=\textwidth}
\caption{Pulsation amplitudes corresponding to the phases shown in 
Fig.~\ref{fig07}. \label{fig08} }
\end{figure}

To resolve frequencies from individual runs and daily data sets,
we selected frequencies from Table~\ref{tab05} that were isolated by at least
$30\,\mu$Hz which included $f1$ through $f6$, $f44$, $f45$, $f48$, $f58$
through $f61$, $f67$ and $f68$. 
We then attempted to fit amplitudes and phases
for each of the frequencies for all of the data subsets. As expected,
most of the fits failed as the pulsation amplitudes were well below the detection
limits. Only for five frequencies were we able to fit phases and amplitudes
to some of the shorter data sets. From the 48 data subsets, we
fit 52 of a possible 480 phases and amplitudes for the five frequencies.
Phases are determined as the time of first maximum amplitude after BJD=2452871.5 and converted to fractional 
phases by dividing by the period. In this manner, an error of 0.1 represents a 10\% change in phase.
The fractional phases and amplitudes are shown in Figs.~\ref{fig07} and \ref{fig08} (and
provided in tabular form in the on-line supporting materials). Horizontal bars 
indicate the time span of the data for
the daily and Group sets. Note that some of the amplitudes are just below the
$4\sigma$ detection limit. We include them because the peaks were fairly 
obvious in the FT and since the frequency is known, a lower detection limit
is not unreasonable.

\begin{table*}
\centering
\caption{Properties of pulsation phases and amplitudes for frequencies 
separated by $>30\mu$Hz. Pulsation amplitudes are compared in a number
of ways include by standard deviations ($\sigma$), average ($\langle\,\rangle$)
and maximum ($_{max}$) amplitudes for various grouping of data. The number
of possible detections is in parentheses next to the category (All, Individual,
Daily, or Groups)\label{tab07b} }
\begin{tabular}{|l|c|c|c|c|c|}\hline
 & $f1$ & $f2$ & $f44$ & $f60$ & $f67$ \\ \hline
\multicolumn{6}{c}{All (54)}\\
\#$_{det}$ & 7 & 16 & 8 & 14 & 16 \\
$\langle A\rangle$ & 0.79 & 1.07 & 1.43 & 1.33 & 1.04 \\ 
$\sigma_A$ &         0.19 & 0.27 & 0.45 & 0.60 & 0.45 \\
$\sigma_A/\langle A\rangle$ & 0.24 & 0.25 & 0.31 & 0.45 & 0.43 \\
$A_{GI-II}/\langle A\rangle $ & 0.78 & 0.64 & 0.43 & 0.35 & 0.42 \\
$\sigma_{\phi}$ (\%) & 4.3 & 12.4 & 19.7 & 17.8 & 17.1 \\
\multicolumn{6}{c}{Individual runs (37)}\\
\#$_{det}$ & 0 & 8 & 4 & 10 & 9 \\
$\langle A\rangle$ & .. & 1.11 &1.72& 1.47 & 1.20 \\
$\sigma_A$ &         .. & 0.29 &0.45& 0.64 & 0.51 \\
$\sigma_A/\langle A\rangle$ &..& 0.26 & 0.26 & 0.44 & 0.43 \\
$A_{max}$ & .. & 1.67 &2.35& 2.75 & 2.41 \\
$\langle A\rangle / A_{max}$ & .. & 0.66 & 0.73 & 0.53 & 0.50 \\
$A_{GI-II}/\langle A\rangle $ & .. & 0.62 &0.35& 0.31 & 0.37\\
$\langle A_{d}\rangle /\langle A\rangle$ & .. & 1.05 &0.65& 0.76 & 0.87\\
$\langle A_{GIII-VI}\rangle /\langle A\rangle$ & .. & 0.79 &0.72& 0.41 & 0.59\\
$A_{GI-II}/A_{max}$ & .. & 0.41 &0.26& 0.17 & 0.18\\
$\langle A_{d}\rangle /A_{max}$ & .. & 0.70 &0.47& 0.40 & 0.43\\
$\langle A_{GIII-VI}\rangle /A_{max}$ & .. & 0.53 &0.53& 0.22 & 0.29\\
$\sigma_{\phi}$ (\%) & .. & 14.6 &26.0& 19.2 & 13.6 \\
\multicolumn{6}{c}{Daily runs (13)}\\
\#$_{det}$ & 3 & 4 & 3 & 3 & 3 \\
$\langle A\rangle$ & 0.90 & 1.17 & 1.11 & 1.11 & 1.04 \\
$\sigma_A$ &         0.05 & 0.13 & 0.26 & 0.25 & 0.13 \\
$\sigma_A/\langle A\rangle$ & 0.06 & 0.11 & 0.23 & 0.23 &0.13 \\
$A_{max}$ & 0.95 & 1.28 & 1.35 & 1.40 & 1.17 \\
$\langle A\rangle / A_{max}$ & 0.95 & 0.91 & 0.82 & 0.79 & 0.89 \\
$A_{GI-II}/\langle A\rangle $ & 0.69 & 0.59& 0.55 & 0.41 & 0.42\\
$\langle A_{GIII-VI}\rangle /\langle A\rangle$ & 0.79 & 0.75 & 1.12 & 0.55 & 0.68\\
$A_{GI-II}/A_{max}$ & 0.65 & 0.54 & 0.45 & 0.33 & 0.38\\
$\langle A_{GIII-VI}\rangle /A_{max}$ & 0.75 & 0.69 & 0.92 & 0.44 & 0.61\\
$\sigma_{\phi}$ (\%) & 4.0 & 14.5 & 8.0 & 15.2 & 12.2 \\
\multicolumn{6}{c}{Groups III - VI}\\
\#$_{det}$ & 4 & 4 & 1 & 1 & 4 \\
$\langle A\rangle$ & 0.71 & 0.88 & 1.24 & 0.61 & 0.71 \\
$\sigma_A$ &         0.22 & 0.29 & .. & .. & 0.33 \\
$\sigma_A/\langle A\rangle$ & 0.31 & 0.33&..&..&0.46 \\
$A_{max}$ & 1.01 & 1.28 & 1.24 & 0.61 & 1.06 \\ 
$\langle A\rangle / A_{max}$ & 0.70 & 0.69 & .. & .. & 0.67 \\
$A_{GI-II}/\langle A\rangle $ & 0.87 & 0.78 & 0.49 & 0.75 & 0.62\\
$A_{GI-II}/A_{max}$ & 0.61 & 0.54 & 0.49 & 0.75 & 0.42\\
$\sigma_{\phi}$ (\%) & 4.9 & 3.4 & .. & .. & 14.0 \\ \hline
\end{tabular}
\end{table*}

Table~\ref{tab07b} examines the amplitude and phase properties organized by
time domain.
While $f1$ is not detected in any individual runs, the phase is stable to within
the errorbars. $f2$ shows deviations of 12\% and $f44$, $f60$, and $f67$ have 
deviations all near 18\%. Phase errors for individual measurements are
under 10\% (except for one) with an average of 5.0\%. Phase variations are provided
in Table~\ref{tab07b} labeled as $\sigma_{\phi}$~(\%). While the deviations appear
similar for $f44$, $f60$, and $f67$, $f44$ has one discrepant value while
$f60$, and $f67$ appear phase variable, particularly from the individual
runs (black circles in Fig~\ref{fig07}). However none of the phases appear
randomly distributed nor do they appear bimodal, which would be an indicator
of unresolved frequencies.

The amplitudes show a larger variety. $f1$ and $f2$ are the most stable
(the smallest $\sigma A/\langle A \rangle $), but the amplitudes for $f1$
are so small that it is not detected in any individual runs and only three
of 13 daily runs. As such, it is clear that the number of detections will
significantly affect amplitude stability as smaller, and therefore more
deviant amplitudes will not have been measured. For these five frequencies,
none are detected more than 30\% of the time. Therefore our measure of
amplitude variability, $\sigma A/\langle A \rangle $ should be
considered a lower limit. 
Likewise, the variations for $f44$, $f60$ and $f67$ must be higher then what
we report as they have some relatively high amplitudes yet are not detected
in all runs. For the amplitudes,
temporally nearby runs can have different amplitudes and consistently, longer
time domains have lower amplitudes.
As these frequencies are reasonably separated from others for all of the
timescales considered and their amplitudes do not appear bimodal, it is 
unlikely that beating plays a role in the
amplitude variations. They are most likely intrinsic to the pulsations.

The original goal of this subsection was to determine
the timescale of phase and amplitude variations. 
Indicators of that timescale are the standard deviations, the
ratios between the average and maximum amplitudes, and comparing these
between the sets. For example, if the timescale
of variation is longer then a day, then the average and maximum values
of the daily runs and the individual ones should be similar and should have
$\langle A\rangle / A_{max}$ near one. For $f2$, the average daily 
amplitude is slightly larger then that for the individual runs while
that for the group sets is significantly smaller. This indicates that
the timescale for amplitude variations is near to or longer than 16 hours
but shorter than 72 hours.

\subsection{Stochastic properties}
An indicator for stochastic pulsations in solar-like oscillators is a
$\sigma_A/\langle A\rangle$ ratio near 0.52 (Christensen-Dalsgaard et al. 
2001). Pereira \& Lopes (2005) also derived this ratio and were the 
first to apply
it in testing whether pulsating sdB stars could be stochastically
excited. Their results for the pulsating sdB star PG~1605+072, based on
seven nights of data, indicated that those pulsations were driven rather
than stochastically excited. The ratios for $f60$ and
$f67$ are near to this value. However in solar-like oscillators
the amplitude decay timescale is long compared to the re-excitation 
timescale, and
so this ratio may not be a good indicator for sdBV stars. Other features
that could indicate stochastic oscillations include significant 
amplitude variability, amplitudes that are reduced in
longer-duration data sets (caused by phase variations), and matches 
with simulated stochastic data. 

Just like for
PG~0048+091, we produced Monte-Carlo simulations for stochastic 
oscillations with varying decay and re-excitation timescales appropriate
for the various data sets and groups we have for KPD~1930.
As there are many amplitude ratios to work with from Table~\ref{tab07b}, it
was hoped that tight constraints for amplitude variations could be
deduced by matching the simulations with the observations. However,
such was not the case and the best we could do was produce loose timescales.
The best results for $f1$ indicate short amplitude decay timescales (four to
six hours) and long re-excitation timescales (20 to 30 hours); those for $f2$
indicate medium decay timescales (nine to 15 hours) and medium to long 
re-excitation timescales (12 to 27 hours); those for $f44$ indicate short
decay timescales (four to nine hours) and medium re-excitation timescales (eight
to 16 hours); those for $f60$ indicate short to medium decay timescales (four
to 15 hours) and long re-excitation timescales (25 to 40 hours); and those
for $f67$ indicate short to medium decay timescales (four to 20 hours) and
long re-excitation timescales (15 to 28 hours). 
The resultant timescales are fairly
consistent in that the decays are always short compared to the
re-excitations, but they are not nearly as clear as
for PG~0048+091, certainly owing to the complexity of KPD~1930's
pulsation spectrum. However, the pulsation phases do not 
appear randomly distributed, as would be expected for stochastic 
oscillations. So we are left with some indications that stochastic
processes may be present, and some contrary information against stochastic
oscillations. From these data, we cannot discern between them.

\subsection{Observed multiplets}
From the ellipsoidal variations in KPD~1930, we can strongly infer it to be
tidally locked, which means we also know the rotation period. In
standard spherical harmonics, each
$\ell$ can produce $2\ell +1$ $m$ azimuthal values separated by the
orbital frequency and the Ledoux constant (which is very small for
$p-$modes in sdB stars).,
we can search for multiplets to impose observational constraints on the
mode degrees ($\ell$). Table~\ref{tab08}
lists multiplets detected to splittings of four times the orbital
frequency of 121.7~$\mu$Hz.
As there are many pulsation frequencies related by a
multiple of the rotation/orbital frequency, the order of the frequencies 
is the same as in Table~\ref{tab05} with the numbers in parentheses indicating
the multiple of the rotation/orbital frequency between it and the previous
(smaller) frequency. The leftmost frequency is 
just the lowest for each multiplet but has no significance otherwise.


\begin{table*}	
\centering
\caption{Pulsation frequencies split by a multiple of the rotation/orbital
frequency. The pulsation frequencies refer to those in Tables~\ref{tab05}
and \ref{tab06}.
The frequencies are in ascending order
and the number in parentheses indicates the multiple
of the orbital frequency between itself and the previously listed
frequency. Column 1 lists the minimum degree for the multiplet using the
classical interpretation. \label{tab08} }
\begin{tabular}{lll}\hline
$\ell_{\rm min}$ & Des. & Designations of related frequencies \\
1 & f1: & f2 (1), f3 (1) \\
1,1 or 3 & f4: & f6 (1), f8 (1), f17 (2), s72 (1), f34 (1) \\
2 & f5: & f10 (2), f15 (1), f24 (1)\\
2 or 4 & f7: & f27 (3), f37 (2), f41 (2) \\
2 & f9 : & f13 (1), f21 (1), f30 (1), f35 (1)\\
1 & f11: & f28 (2)\\
1 & f12: & f29 (2) \\
3 & f14: & f31 (2), s75 (2), f43 (2)\\
1 & s70: & f33 (2)\\
1 or 3 & f23: & f39 (4), f44 (1) \\
2 & f25: & s73 (2)\\
3 or 5 & s71: & f40 (4), f48 (3), f51 (3)\\
1 & f26: & s74 (2)\\ 
1 & f42: & f46 (2) \\
2 & s76: & s79 (2), f55 (2) \\
4 or 3 & s78: & f50 (2), f56 (2), f59 (1), f65 (3)\\
2 & f49: & f61 (4) \\
1 & f57: & f62 (2) \\
1 & f60: & f64 (1)\\
1 & f63: & f66 (2) \\ \hline

\end{tabular}
\end{table*}

\subsubsection{Classical interpretation}	
In total, 61 of our 81 frequencies are related by a multiple of
the rotation/orbital frequency. In a classical asteroseismological
interpretation, we would assume the spin axis is aligned with the orbital
axis and the stars are tidally locked. As such, we are viewing the spin
axis close to equator-on, or an inclination near $80^o$ and the multiplets
are caused by stellar rotation.

As each degree can have $2\ell +1$ azimuthal orders, $m$, the number of
orbital splittings provides a minimum $\ell$ value and constrains where
the $m=0$ is. For example, the $f1,\,f2,\,f3$ triplet is best interpreted
as an $\ell =1$ triplet with $m=0$ at $f2$. The multiplet beginning with $f4$
has sufficient frequencies to warrant an $\ell =3$ interpretation. However,
$\ell =3$ modes have low amplitudes caused by geometric cancellation, 
making such an interpretation unlikely. More feasible is that there are 
two $\ell =1$ multiplets with their outside components at a chance 
separation of nearly $2f_{orb}$. Column 1 of Table~\ref{tab08} gives the 
minimum $\ell$ degree based on the number of orbital splittings. For entries
with multiple possibilities, the most likely is given first. 
(On-line supporting material includes a color-coded figure 
showing the multiplets
and a figure showing just the $m=$ components with their corresponding
degree.)


We can  also use the pulsation amplitudes to place  some 
constraints on the modal degrees. Using Groups I and II as a guide, all
of the amplitudes are within a factor of 7 of each other, while most are
within a factor of 4. Table~\ref{tab09} lists the geometric cancellation
factors for azimuthal orders $\ell ,\, m\,=\,1,0$ through $4,4$ for 
rotation axes of
$i_r=70$ and $80^o$. These factors indicate how much a pulsation amplitude would
be reduced relative to a radial mode, which suffers no geometric
cancellation. As column 2 indicates, for the classical interpretation, all
of the $\ell \leq 2$ modes have amplitudes reduced by factors less than
5 to 8, depending on orientation. This is roughly in agreement with the 
observed amplitude range. By
contrast, all $\ell \geq 3$ modes have amplitudes
reduced by factors $\geq 20$ (except for $\ell ,\,m=4,|1|$ depending on
viewing angle). Since we don't observe this range of amplitudes
in KPD~1930, it is unlikely that these multiplets are 
being observed.

\begin{table}  
\centering
\caption{Geometric cancellation (pulsation amplitude reduction)
factors for $i_r=80^o$ ($i_r=70^o$ in parentheses).
\label{tab09} }
\begin{tabular}{lll}\hline
$\ell\,,m$ & Classical & Tipped \\
$1,0$ & 4.81 (2.44) & 1.69 (1.77) \\
$1,|1|$ & 1.20 (1.26) & 2.40 (2.52) \\
$2,0$ & 3.35 (4.70) & 4.18 (4.59) \\
$2,|1|$ & 7.28 (3.87) & 5.15 (5.65) \\
$2,|2|$ & 2.57 (2.82) & 5.48 (7.67) \\
$3,0$ & 48.2 (28.9) & 39.9 (45.9) \\
$3,|1|$ & 32.9 (70.6) & 860 (72.6) \\
$3,|2|$ & 51.7 (28.9) & 73.0 (57.8) \\
$3,|3|$ & 22.3 (25.7) & 68.1 (51.8) \\
$4,0$ & 32.9 (7,911) & 34.6 (41.8) \\
$4,|1|$ & 41.7 (4.81) & 7.46 (8.04) \\
$4,|2|$ & 30.0 (151) & 55.0 (45.9) \\
$4,|3|$ & 173 (21.4) & 64.2 (57.2) \\
$4,|4|$ & 81.4 (98.1) & 282 (218) \\ \hline

\end{tabular}
\end{table}

Ignoring high-degree modes, there remains multiplets as evidence of
12 $\ell =1$ and four $\ell =2$ modes. There are also 20 frequencies which
show \emph{no} relations to other frequencies via an orbital overtone. These
are all candidates for radial modes. As there is no geometric cancellation
for radial modes, they would be expected to have some of the higher
amplitudes. $f16,\,f18,\,f19,$ and $f20$ all have amplitudes greater than
1~mma for all of their detections. Yet none of these
have amplitudes significantly higher than the other frequencies. Therefore
it is also possible they are $\ell > 0$ modes with only one frequency
visible.

\subsubsection{Tipped-axis interpretation}
Another interpretation would be that the pulsation axis is aligned
with the tidal force of the companion. As described by Reed, Brondel,
\& Kawaler (2005), such a pulsation geometry would precess, completing a
revolution every orbital period. The change in viewing pulsation geometry
incorporates three additional pieces of information into the lightcurve
which can be used to uniquely identify the pulsation degree $\ell$ and the
absolute value of the azimuthal order $|m|$. These are: a pattern of peaks
in the FT of the integrated lightcurve; two or more $180^o$ flips in pulsation 
phase over an orbital period; and recovery of the ``true'' peak in the FT
of the phase-separated data. As in Reed et al. (2005), we will not seek
analytic solutions, but will use simulated data of precessing pulsation
geometries to guide us. These simulations and any tipped-axis analysis make
the assumption that spherical harmonics apply.

\begin{figure}
 \centerline{
\psfig{figure=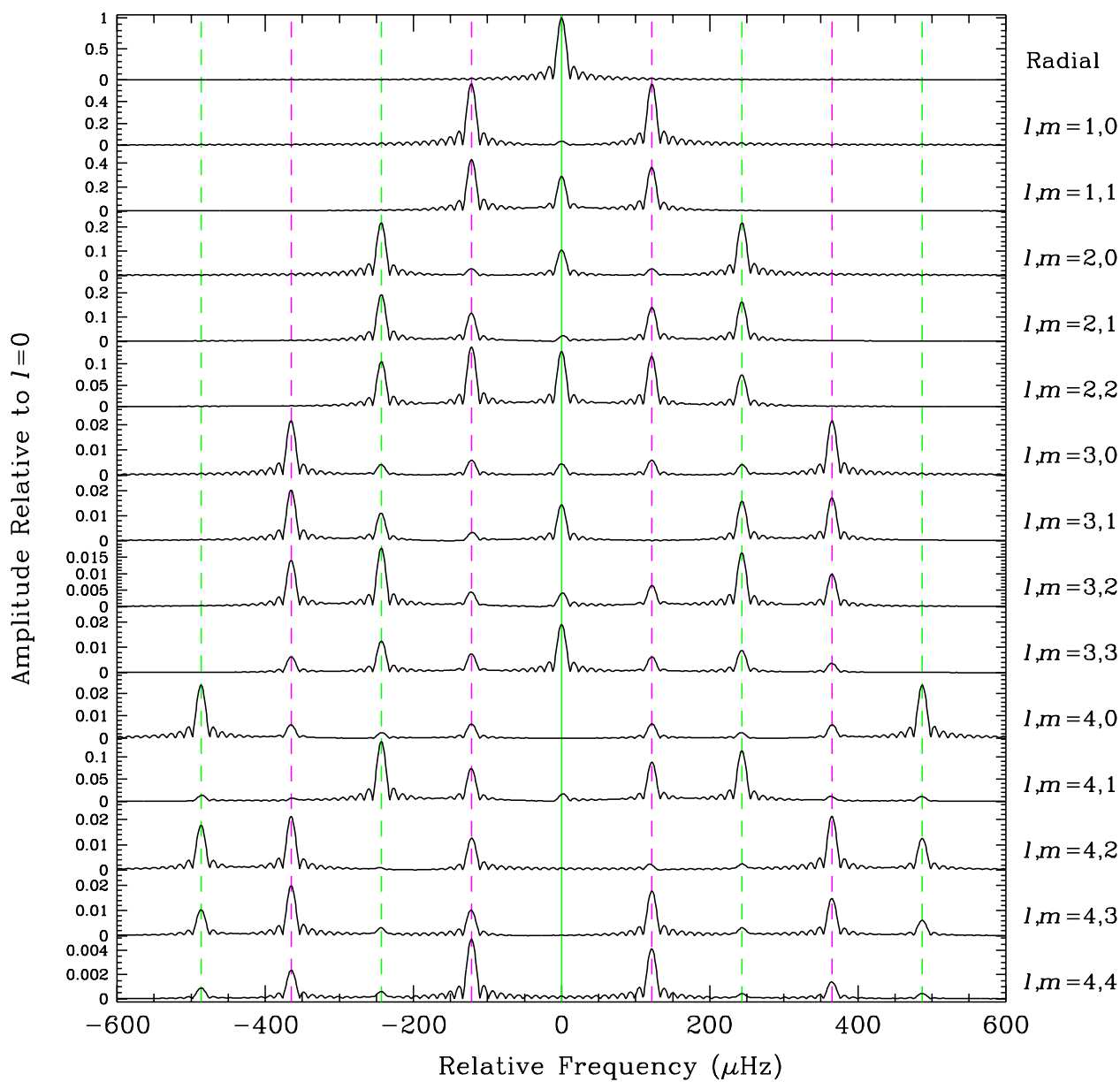,width=\textwidth}}
 \caption{Pulsation spectra of simulated data where the pulsation pole
points $5^o$ off the orbital axis, which has an inclination of $70^o$. 
The input frequency is
at the center (solid green line) and the dashed lines indicate orbital
aliases. The amplitudes are relative to the radial mode and the mode is 
indicated on the right.
\label{fig14} }
\end{figure}

We begin by searching for patterns in the FT of the integrated data of 
KPD~1930. To guide our search, we produced simulated data
for modes from $\ell =0$ through $4$ for an orbital/rotation axis of
$i_r=70^o$ and a pulsation axis of $i_p=85^o$ relative to the rotation
axis. This geometry seems reasonable for what we know of the orbital
inclination and with tidal forces larger than the maximum Coriolis
force. Changes of five to ten degrees in
either axis make little difference in the patterns. The simulated FTs are
shown in Fig.~\ref{fig14}. The amplitudes are
relative to the $\ell =0$ (radial) mode and along with the geometric
cancellation factors of Table~\ref{tab09} indicate that high-degree
modes are unlikely to be observed. The frequency patterns of Fig.~\ref{fig14}
are what we are looking for in the observed data.

Eighteen of the multiplets from Table~\ref{tab05} match patterns from
simulated tipped-axis pulsations\footnote{These are shown schematically
in Fig.~2 of the on-line supporting materials.}. The next step is to
divide the data into orbital regions of like and opposing pulsation phases
(Reed et al 2005 show this procedure in detail) for the different modes. 
If the intrinsic 
frequencies predicted by the tipped-axis model are recovered, and their 
phases are opposite
in appropriate data sets, this would be a reasonable indicator of a tipped 
pulsation axis. As can be seen in Fig.~\ref{fig14},
the intrinsic frequency for tipped modes $\ell\, m=1,1;\,2,0;\,
2,2;\,3,1$ and $3,3$ will have a corresponding frequency
 in the combined data. As such, recovering the peak in the divided data is
less important but detecting a phase shift  of 0.5 will
be vitally important for associating those frequency patterns with tipped
pulsation modes.


We separated the data for Groups I through V
into orbital regions of like and opposing phases appropriate for modes
$\ell ,\,m\,=\,1,0$ through $4,4$ and searched each Group 
not only for
the frequencies predicted from the observed multiplets, but for any previously
unobserved frequencies above the noise\footnote{The phased data sets for 
Group II with $\ell ,\,m\,=\,1,0$ and $1,1$ are shown in Fig.~5 of the 
on-line supporting materials.}.
The simplest data sets are those for $\ell =1$
as the orbit is divided into halves,
with set A of $\ell ,\,m\,=\,1,1$ going from an orbital 
phase of 0.0 to 0.5 and set B covering the other half. Those for 
$\ell ,\,m\,=\,1,0$ are shifted by $-0.25$ in phase. 
Table~\ref{tab10} provides the results of the search. Column 1 provides a
unique mode identifier (with a corresponding identifier from Table~\ref{tab05}
in parentheses, if there was one), 
column 2 the frequency, column 3 indicates the mode
the data was phase-separated for and the remaining columns give the phase 
difference (set A minus set B) for each group.


\begin{table}
\centering
\caption{Phase differences (A-B) for possible tipped pulsation frequencies.
NLLS errors are in parentheses.
Differences near $\pm 0.5$ indicate tippled-axis modes.
\label{tab10} }
\begin{tabular}{llcccccc}\hline
ID & Freq. & MODE & GI & GII & GIII & GIV & GV \\
$t1 (f17)$ & 3913 & 1,0 & - & - & - & - & - \\
$t2 (f26)$ & 4008 & 1,0 & -0.50 (2) & 0.82 (2) & - & - & - \\
$t3 (f32) $& 4113 & 1,0 & - & - & - & - & - \\
$t4 (f33)$ & 4126 & 1,0 & - & - & - & - & - \\
$t5$ & 4647 & 1,0 & 0.55 (6)& -0.45 (5)& 0.27 (4) & -0.43 (6)& - \\
$t6$ & 5581 & 1,0 & - & - & - & - & - \\
$t7$ & 5831 & 1,0 & - & - & - & - & - \\
$t8 (f2)$ & 3187 & 1,1 & -0.55 (3) & 0.49 (5) & - & 0.46 (4)& - \\
$t9 (f6)$ & 3543 & 1,1 & -0.67 (3)& -0.64 (3)& -0.51 (4)& 0.39 (4)& 0.41 (4) \\
$t10 (f15)$ & 3853 & 1,1 & -0.08 (3)& -0.07 (2)& -0.16 (3)& -0.01 (2)& -0.13 (3) \\
$t11 (f28/29)$ & 4030 & 1,1 & -0.25 (2)& -0.26 (2)& - & -0.39 (2)& -0.26 (3)\\
$t12 (f37)$ & 4263 & 2,0 & - & 0.51 (2)& - & -0.31 (4)& - \\
$t13$ & 5140 & 2,0 & - & - & - & - & - \\
$t14 (f10)$ & 3731 & 2,2 & -0.22 (2)& - & - & -0.27 (2)& - \\
$t15 (f21)$ & 3926 & 2,2 & - & - & - & - & - \\
$t16 (f50)$ & 5207 & 3,1 & - & - & - & - & - \\
$t17$ & 4210 & 3,2 & -0.54 (5)& - & - & -0.63 (3)& - \\
$t18 $ & 4188 & 4,3 & - & - & - & - & - \\
$t19 $ & 4188 & 4,4 & 0.61 (4)& - & -0.47 (5)& - & - \\ \hline

\end{tabular}
\end{table}

As anticipated considering the complexity of the data, the results are
not straightforward. In the split data sets the aliasing is significantly
worse and the pulsation amplitudes are very low. Evidence also suggests 
that amplitudes
and phases are changing throughout the campaign.
%
Most of the predicted frequencies are not detected. Those that are detected
have amplitudes only marginally above the noise. But
the strongest evidence will be consistent phase differences of one half
in the various data sets.
Frequency $t10\,(f15)$ shows no phase shift between sets for all 
groups and so cannot be a tipped pulsation mode. Frequencies $t11$ and $t14$
have intermediate values, again indicating that they are likely not
tipped pulsation modes, but rather that their phases are not stable over
the course of the observations. Frequencies $t2$ and $t12$ are only detected 
in two of
the five data sets, and while one phase difference is near one half, the
other has an intermediate value. As $t2$ is also seen in the integrated 
lightcurve, but shouldn't be for an $\ell ,m\,=\,1,0$ mode, it is unlikely
a tipped mode.
 Frequencies $t5,\, t8$, and  $t9$ do
fit what we expect for tipped pulsations. $t9$ is detected with a similar
phase difference in all five groups, $t8$ has
consistent phase differences for the three data sets in which it can be
detected, and $t5$
has consistent phase differences in three of four detections. Additionally,
$t5$ is not detected in the integrated data, as should be the case. 
More surprising are the results for $t17$ and $t19$,
both of which are only detected in two of the five groups, but have phase
differences near to one half. These modes are unlikely to be
observed because the geometric cancellation factors are very high
(73 and 282, respectively), meaning
the intrinsic amplitudes would need to be much higher than the others.
Monte Carlo simulations were produced with random phases between data sets. 
This could be appropriate for purely stochastic pulsations, but for driven
pulsations, the phases should not be random at all, but rather close to a fixed
number. As tipped pulsations have a phase shift of 0.5 which is very 
unexpected for driven pulsations, the significance of the simulations are
somewhat startling. Our phase detections for $t8$ and $t9$ only occur
in 0.1\% of our simulations while those for $t5$ occur 1.0\% of the time.
Those for $t17$ and $t19$ occur 6.7 and 4.9\% of the time, repectively.

Of the 18 frequency multiplets in the combined data sets, three possess
indicators for tipped pulsation modes. Nine have no detections at all, two
have phase differences near zero, and two more have marginally appropriate
phase differences, but would indicate modes that are unlikely to be
detected because of geometric cancellation. Still, particularly for $t5$,
which is not observed in the integrated data, the phase difference is
a precise indicator for tipped pulsation modes. The chance of this occurring
randomly is quite small.

\subsection{Orbital dependence}
\begin{figure}
 \centerline{
\psfig{figure=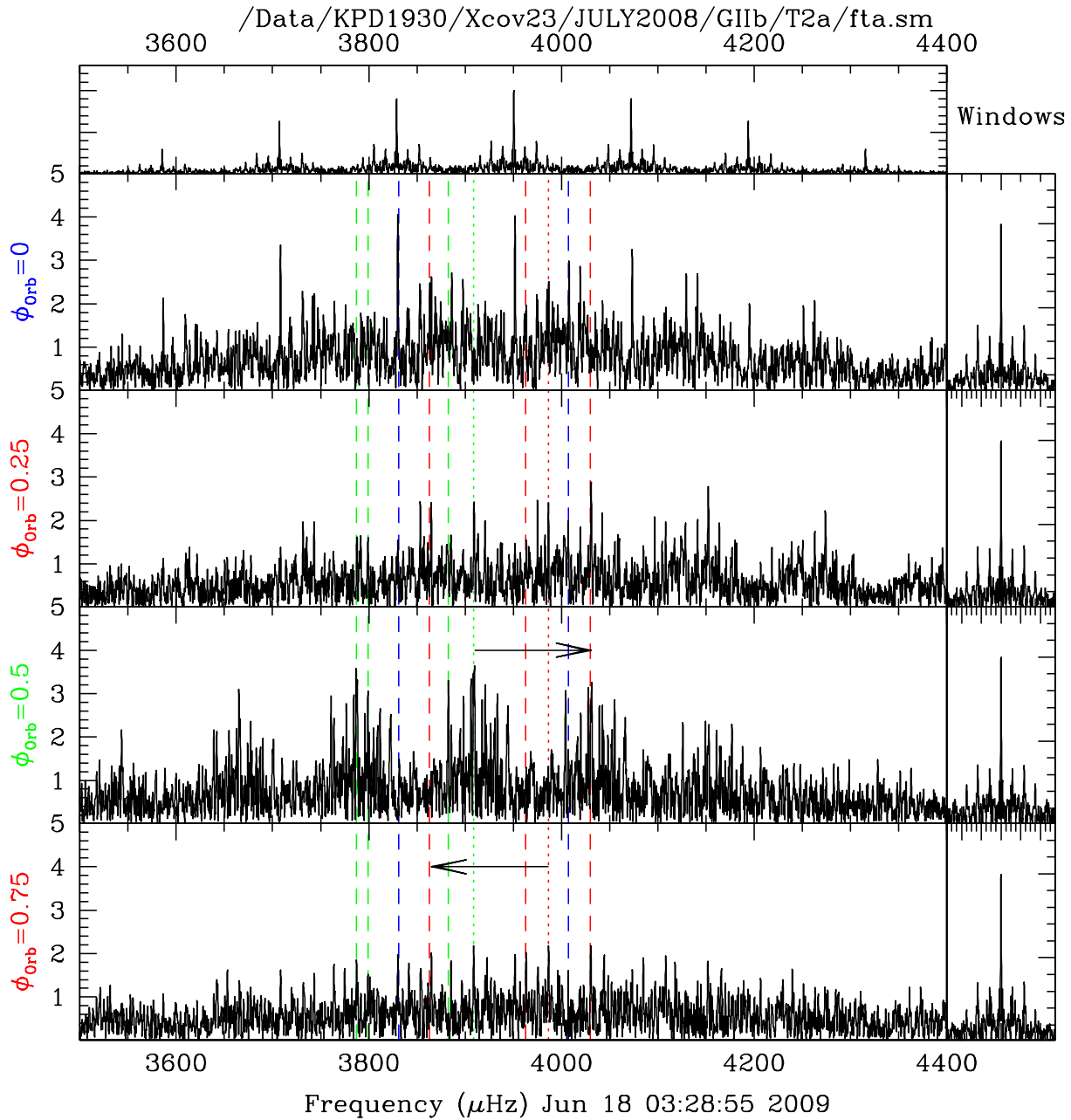,width=\textwidth}}
 \caption{Pulsation spectra of Group II's data, separated into four
subsets based on orbital phase. Dashed lines indicate frequencies which show
an orbital dependence and dotted lines indicate frequencies an orbital alias
away from the most likely frequency (indicated with an arrow and color-coded
in the on-line version).  Top and right panels are window functions.
\label{fig17} }
\end{figure}

While searching for tipped-axis pulsations, we stumbled upon the unexpected
finding that some frequencies appeared to only occur for some stellar
orientations\footnote{These can be seen in Fig.~5 of the supporting
materials.}.
So we divided the data up into four, eight, and 10 subsets 
according to orbital phase and examined the pulsation spectra of each.
We found that quadrants, centered around the orbital phases 0 (QI), 0.25 (QII),
0.5 (QIII), and 0.75 (QIV) sufficiently differentiated the orbital dependence
of the frequencies. Temporal spectra of
Group II's data, split into four quadrants are shown in Fig.~\ref{fig17}.
The dashed lines indicate likely intrinsic frequencies while the dotted
lines are more likely aliases, with the arrows indicating which frequency
they are an alias of.

It is obvious that there is an orbital dependence on some frequencies, 
particularly $o3$, which is clearly and only seen in QI. This
most likely means that the pulsation geometry, as normally described 
by spherical harmonics, is more complex for these frequencies. 
Describing the pulsation geometry for such modes is beyond the scope
of this paper, but we provide frequencies which have a detectable orbital
dependence in Table~\ref{tab11}. Outside of the main area of power, there is
little sign of this uniqueness other then $f57$ (5459) has a slightly higher
amplitude in quadrant $\phi\,=\,0.25$ and $f67$ (6319) is higher 
in quadrants $\phi\,=\,0$ and $0.5$. Undoubtedly the lack of other detections
is caused by low amplitudes combined with severe aliasing.

\begin{table}
\centering
\caption{Pulsation frequencies that have an orbital dependence. o- and o+
indicate frequencies detected an orbital alias away while d+ indicates a
frequency a daily alias away.
\label{tab11} }
\begin{tabular}{llcccc}\hline
ID & Freq. & QI & QII & QIII & QIV  \\
$o1$ & 3787 &  &  & X & \\
$o2$ & 3799 & & & X & \\
$o3$ & 3831 & X  & &  & \\
$o4\,(f16)$ & 3865 &   & X&  & o-\\
$o5$ & 3882 & & & X & \\
$o6\,(f23)$ & 3963 & & d+ & & X \\
$o7$ & 4008 & X &  &  &  \\ 
$o8$ & 4031 &  & X & o- & X \\ \hline

\end{tabular}
\end{table}


\section{Results} 
We have analyzed WET data spanning 26 days in August and September 2003,
supplemented by 45 hours of multisite data obtained in July 2002. These
data were used to examine the orbital properties and pulsation spectrum
of the sdB+WD binary KPD~1930.
We used the ellipsoidal variation to affirm the orbital period and
folded the data over that period. No signs of eclipse were detected,
constraining the inclination to $<78.5^o$. Additionally, we noted
that the minima are uneven, indicating that KPD~1930 is very
slightly asymmetric and marginally detect anticipated Doppler
effects from possibly uneven maxima. 

As implied from the discovery data (B00), we have found KPD~1930 to be
an extremely complex pulsator with frequencies and amplitudes that can vary
on a daily timescale. In these data, we confidently detect 68 pulsation
frequencies and suggest a further 13.
Of these, only 26 are related to frequencies observed by B00; a surprisingly
small number that attests to KPD~1930's pulsational complexity. Our
WET data, which cover more than three weeks during 2003, 
have over four times better temporal resolution and one half the 
detection limit of B00.

We examined amplitude and phase stability by analyzing 
subsets of data over several time scales for well-separated frequencies. 
Unfortunately the low
amplitude of these frequencies hampered our investigation and we were unable
to detect them in most of the subsets. For the times we could detect them,
we found the pulsation
amplitudes to be fairly variable, although all $\sigma_A / \langle A \rangle$
ratios are short of the 0.52 used in solar-like oscillators to indicate
stochastic oscillations (Christensen-Dalsgaard et al. 2001). 
The low ratio could be an artefact of only a few amplitude detections or
caused by an amplitude decay timescale shorter
then the re-excitation timescale. To investigate timescales 
we compared the observed amplitude
ratios for several subsets with simulated stochastic oscillations. The 
simulations could easily fit the observed ratios and broadly found
decay timescales near 12 hours and re-excitation timescales near
25 hours. As such, the pulsations do have some qualities normally associated
with stochastic oscillations, however the phases are relatively stable which
argues against stochastic oscillations. As such, it is possible that the
amplitude variations have a different cause.

\begin{figure}
 \centerline{
\psfig{figure=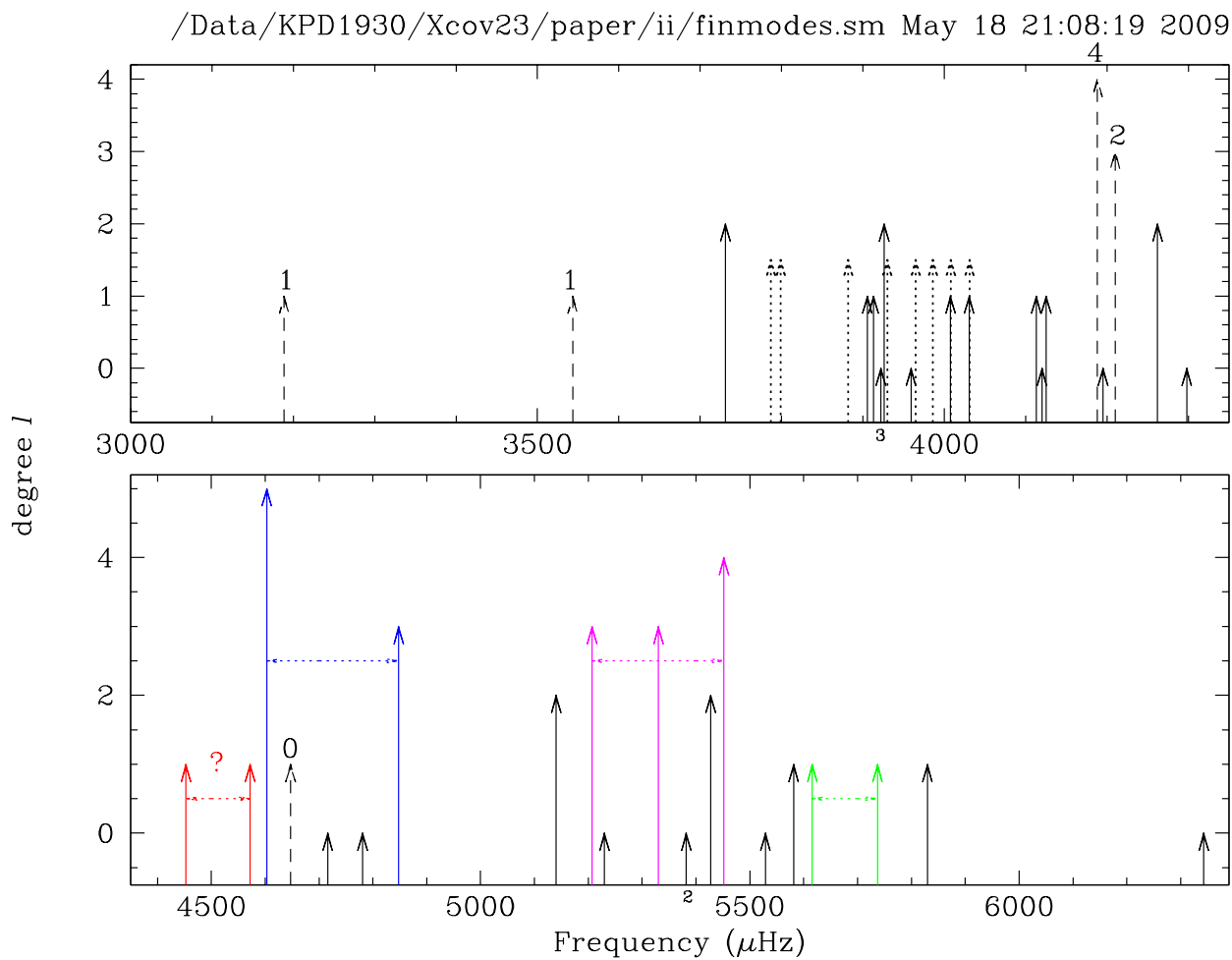,angle=-90,width=\textwidth}}
  \caption{Schematic of modes for KPD~1930. The modal degree ($\ell$) 
is indicated by arrow
length.  Solid lines indicate the 
$m=0$ component of classically
interpreted multiplets.
For those multiplets with more then one possible $m=0$
frequency, they are connected by a dotted horizontal line.
(They are also color-coded in the electronic version.) The dashed lines
indicate tipped-axis modes, with the $m$ index above the arrow and the 
dotted lines indicate frequencies which show a dependence on orbital phase
(set to an arbitrary amplitude of 1.5, as no degree was ascertained for these
frequencies).
\label{fig18} }
\end{figure}

We have found 20 separate multiplets including up to 61 frequencies.
Assuming the classical interpretation which aligns the pulsation
axis with the rotation axis, these would indicate 12 $\ell =1$,
four $\ell =2$, and/or possibly (but unlikely) four $\ell = 3$, two $\ell =4$,
and one $\ell =5$ modes. We also searched for indicators of a tidally-induced
tipped pulsation axis which would precess with each orbit. While the
complexities of the pulsations made searching for tipped modes difficult,
three frequencies were found which indicated tipped modes and two further
modes had some indications of tipped-modes, but were for $\ell =3$ and $4$,
which are unlikely to be observed. We also detected
frequencies which only appeared during certain orbital phases, which
indicates that some frequencies are affected by the slight 
asymmetry of the star.
Figure~\ref{fig18} schematically shows these results with the arrow height
indicating degree (except as noted below). Classically interpreted
multiplet $m =0$ components are shown as solid arrows (with multiple
possible $m=0$ frequencies connected by a horizontal line as in 
Fig.~\ref{fig14} and color-coded in the electronic version); 
modes from the tipped
axis interpretation are shown as dashed arrows, with the $m$ value above
the arrow; frequencies which only appear during certain orbital phases
are shown as dotted arrows (at an arbitrary height of 1.5 since we have
no indication of their degree). Any 
frequencies that were
not involved in the above cases were (somewhat arbitrarily) deemed to be
$\ell =0$ modes and are indicated as such in the figure with solid arrows.
Note that the classically-interpreted $\ell =1$ mode which could have its
$m=0$ component at 4453 or 4572 is marked with a question mark. This is
part of the tipped $l,\,m\,=\,3,\,2$ multiplet and so \emph{if} the tipped
mode is correct, then the classical mode would be invalid.
The small ``3'' and ``2'' under the axes are to indicate that there
are three and two closely-spaced frequencies, all of which fit the conditions
we ascribe to $\ell =0$ modes that would be unresolvable in the figure.

In the end, KPD~1930 is a star that incorporates a little bit of everything;
many pulsation frequencies, multiplets, indications of a tidally-induced 
pulsation geometry, non-sphericity, relativistic Doppler effects, 
amplitude variations, and ellipsoidal variations, all wrapped up in a 
likely pre-Type~Ia supernova binary. Unfortunately, all these effects complicate
the analysis, making it a bit of an unruly mess. While we have tried to
unravel it, our results indicate that even the WET fails to fully resolve
the complexity of the pulsation spectrum and we can only imagine that
perhaps MOST or Kepler-like data\footnote{For information see
http://www.astro.ubc.ca/MOST/ and http://astro.phys.au.dk/KASC/}
 would be required to improve this
situation. Unfortunately, the Kepler field just misses KPD~1930.
KPD~1930 remains a fascinating star that warrants further 
attempts to understand its complexities.

\section*{Acknowledgments}

This material is based upon work supported
by the National Science Foundation under Grant No. 0307480.
Any opinions, findings, and conclusions or recommendations expressed
in this material are those of the author(s) and do not necessarily
reflect the views of the National Science Foundation.
This research was supported in part by NASA through the American 
Astronomical Society's Small Research Grant Program. MDR was partially
funded by a Summer Faculty Fellowship provided by Missouri State
University, would like to acknowledge HELAS travel funding, and thank
Dr. Conny Aerts for helpful discussions. 
MAM, SLH, SP, JRE, and AMQ were suported by the Missouri
Space Grant Consortium, funded by NASA. 
We would also like to thank several other observers who
assisted with observing, including E. Brassfield, D. McLemore,
J. Peacock, R. Knight, T. Kawasaki, E. Hart, N. Purves, A. Nishimura,
M. Hyogo, and E. Rau.

\section{Supporting Information}
\begin{itemize}
\item Figure~S1: Detailed prewhitening sequence of Group~IV data between
3~600 and 4~250~$\mu$Hz.
\item Table~S1: Pulsation phases and amplitudes for frequencies
separated by $>30\mu$Hz for individual runs, daily combined runs,
and Groups of data.
\item Figure~S2: Schemitic of frequency spacings commensurate with the orbital
frequency.
\item Figure~S3: Schematic associating pulsation frequencies with 
modes for 
traditionally-interpreted multiplets.
\item Figure~S4: Schematic associating pulsation frequencies with possible
tipped modes.
\item Figure~S5: Pulsation spectra of Group II's data, separated into
opposing phases appropriate for $\ell ,\,m=1,0$
and $1,1$ tipped pulsation axis modes.
\end{itemize}

\input{supp_astroPH.tex}

\label{lastpage}

\end{document}

%% file: supp_astroPH.tex
\section{On-line Only Material}

\begin{figure*}
 \centerline{
\epsfig{figure=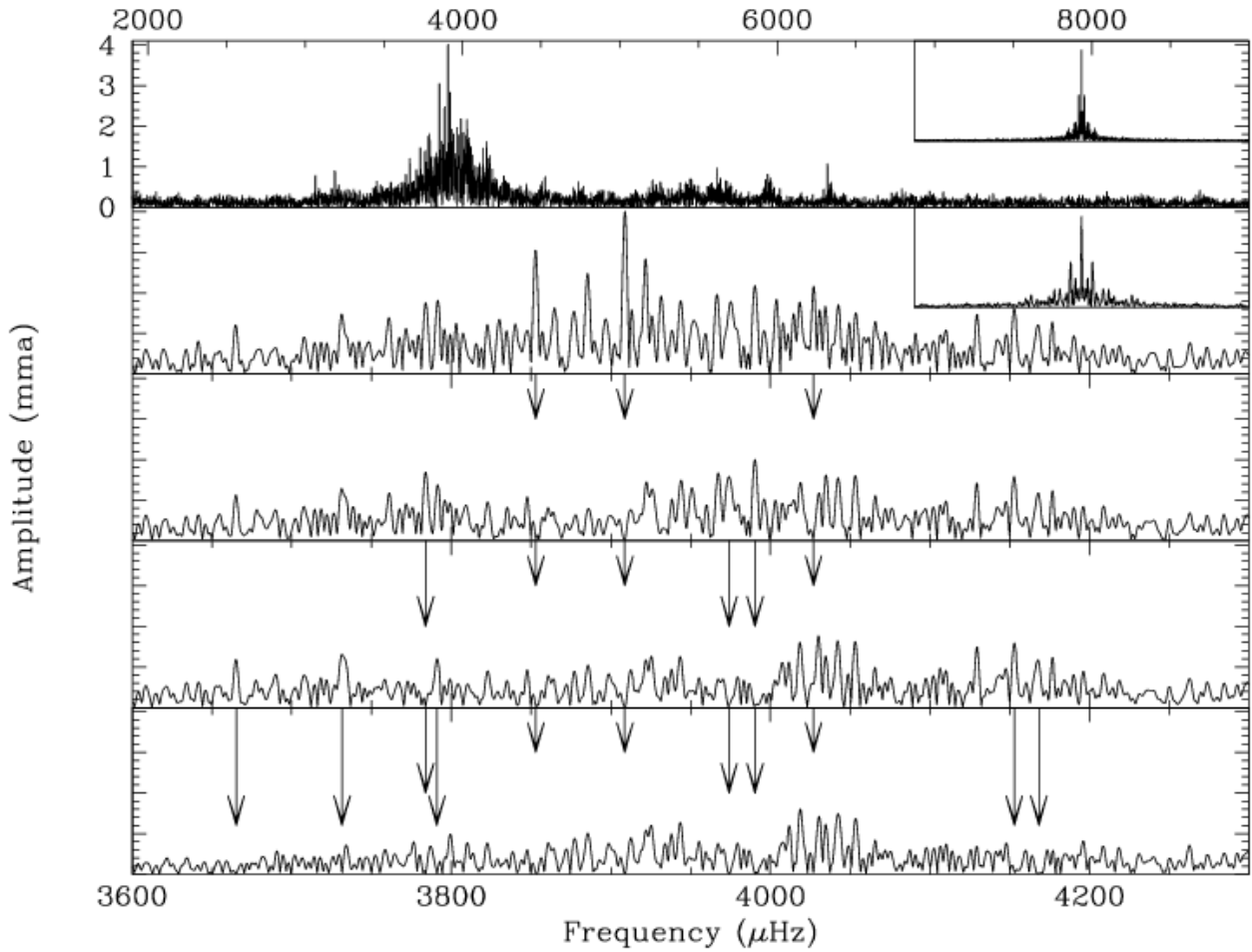,width=\textwidth}}
  \caption{Sequence of least-squares fitting and prewhitening of the Subset
IV data set. The top two panels show the original FT with the top one showing
the FT over a large range, and the second one enlarged to show just a
700~$\mu$Hz region. The following three panels show the original FT
fit and prewhitened by the highest amplitude 3, 6, and 11 peaks,
respectively (indicated by the arrows). The region around 4~000~$\mu$Hz cannot
be successfully prewhitened.
The insets in the top two panels are the window function
plotted at the appropriate horizontal scales. Note that the vertical scale does
not change between panels. \label{fig04} }
\end{figure*}
\pagebreak

\begin{table*}
\centering
\caption{Pulsation phases and amplitudes for frequencies separated by
$>30\mu$Hz for individual runs, daily combined runs, and Groups of data. 
The last column provides the $4\sigma$ amplitude detection limit.
The bottom two rows give combined standard deviations and the stochastic
parameter for amplitudes.
Formal least-squares errors are provided in parentheses.
\label{tab07} }
\begin{tabular}{|l|c|c|c|c|c|c|c|c|c|c|c|}\hline
 & \multicolumn{5}{c}{Phase ($T_{\rm max}/P$)} & \multicolumn{5}{c}{Amplitude (mma)}\\
Run: & $f1$ & $f2$ & $f44$ & $f60$ &  $f67$ & $f1$ & $f2$ & $f44$ & $f60$  & $f67$ & $4\sigma$\\ \hline
not20aug        & .. &0.81 (6)& .. &-0.23 (7)&  .. & .. &0.8 (3)& .. &0.8(3)&  .. &0.80 \\
not21aug        & .. &0.87 (4)& .. & .. &  .. & .. &1.1 (3)& .. & .. &  ..  &0.79\\
not22aug        & .. &0.77 (6)&0.28 (4)& .. &  .. & .. &0.9 (3)&1.3 (3)& .. &  ..  &0.81\\
not23aug        & .. & .. & .. & .. &  .. & .. & .. & .. & .. & ..  &0.82\\
lulin27aug      & .. &0.98 (7)& .. & .. & 0.18 (7)& .. &1.0 (4)& .. & .. & 1.1 (4) &1.02\\
a0688           & .. & .. & .. &0.11 (3)&  .. & .. & .. & .. &2.8 (5)&  ..  &1.48\\
lulin28aug      & .. & .. & .. & .. & 0.23 (9)& .. & .. & .. & .. & 0.9 (5) &1.20\\
gv30809         & .. & .. & .. & .. & 0.31 (4)& .. & .. & .. & .. &1.4 (4) &1.30\\
turkaug29       & .. & .. & .. & .. & 0.34 (9)& .. & .. & .. & .. & 2.4 (1.4) &1.98\\
saao30aug       & .. &0.88 (4)& .. & .. &  .. & .. &1.7 (5)& .. & .. & ..  &1.81\\
mdr248          & .. &0.90 (7)& .. &-0.19 (4)&  .. & .. &1.0 (3)& .. &1.3 (3)&  ..  &1.01\\
haw31aug        & .. & .. & .. &-0.14 (5)&  .. & .. & .. & .. &2.1 (7)&  ..  &1.81\\
turkaug31       & .. & .. & .. &-0.13 (8)&  .. & .. & .. & .. &1.6 (8)&  ..  &1.45\\
turksep01       & .. & .. &-0.11 (9)& .. &  .. & .. & .. &1.5 (8)& .. &  ..  &1.58\\
turksep02       & .. & .. & .. &0.32 (12)& 0.27 (13)& .. & .. & .. &1.2 (9)& 1.2 (9) &1.19\\
wise03sep       & .. & .. &0.43 (6)& .. &  .. & .. & .. &1.6 (5)& .. &  ..  &1.46\\
haw05sep        & .. & .. &0.46 (5)& .. &  .. & .. & .. &2.4 (7)& .. &  ..  &1.96\\
hunsep06        & .. & .. & .. &0.23 (6)&  .. & .. & .. & .. &1.3 (5)&  ..  &1.82\\
wise06sep       & .. &0.68 (5)& .. &0.16 (8)&  .. & .. &1.4 (5)& .. &1.4 (7)&  ..  &1.48\\
hunsep07        & .. & .. & .. &0.01 (5)&  .. & .. & .. & .. &1.8 (6)&  ..  &1.53\\ \hline
aug25           & .. & .. & .. & 0.17 (5)& 0.48 (4)& .. & .. & .. &1.0 (3)& 1.2 (3) &1.41\\
aug27           & .. &1.06 (4)& .. & .. &  .. & .. &1.2 (3)& .. & .. &  ..  &1.22\\
aug28           &0.40 (5)&0.86 (3)&0.52 (5)&-0.02 (7)& 0.24 (4)&0.9 (3)&1.3 (3)&0.8 (3)&1.4 (3)& 1.0 (3) &1.10\\
aug29           &0.42 (5)& .. & .. & .. & 0.33 (4)&1.0 (2)& .. & .. & .. & 0.9 (2) &0.90\\
aug30           &0.34 (5)&0.89 (4)& .. &-0.12 (5)&  .. &0.9 (3)&1.0 (3)& .. &0.9 (3)&  ..  &0.96\\
sep3            & .. & .. &0.43 (5)& .. &  .. & .. & .. &1.1 (4)& .. &  ..  &1.24\\
sep4            & .. & .. &0.36 (4)& .. &  .. & .. & .. &1.4 (4)& .. &  ..  &1.30\\
sep5            & .. &0.70 (5)& .. & .. &  .. & .. &1.2 (4)& .. & .. &  ..  &1.36\\ \hline
Group III       &0.39 (4)&0.83 (4)& .. & .. & -0.11 (7)&0.5 (1)&0.6 (1)& .. & .. & 0.3 (1) &0.44\\
Group IV        &0.37 (3)&0.89 (3)& .. &0.21 (4)& 0.22 (2)&0.7 (1)&0.9 (1)& .. &0.6 (1)&1.1 (1) &0.52\\
Group V         &0.39 (9)&0.81 (3)&0.40 (3)& .. & 0.11 (6)&0.6 (2)&0.8 (1)&1.2 (2)& .. & 0.9 )2) &0.77\\
\hline
$\sigma$        &0.04&0.12&0.20&0.18&0.17&0.19&0.27&0.45&0.60&0.45 \\
$\sigma A/\langle A\rangle$ & & & &    & &0.24&0.25&0.31&0.45&0.43 \\
\hline
\end{tabular}
\end{table*}
\pagebreak

\begin{figure}
\epsfig{figure=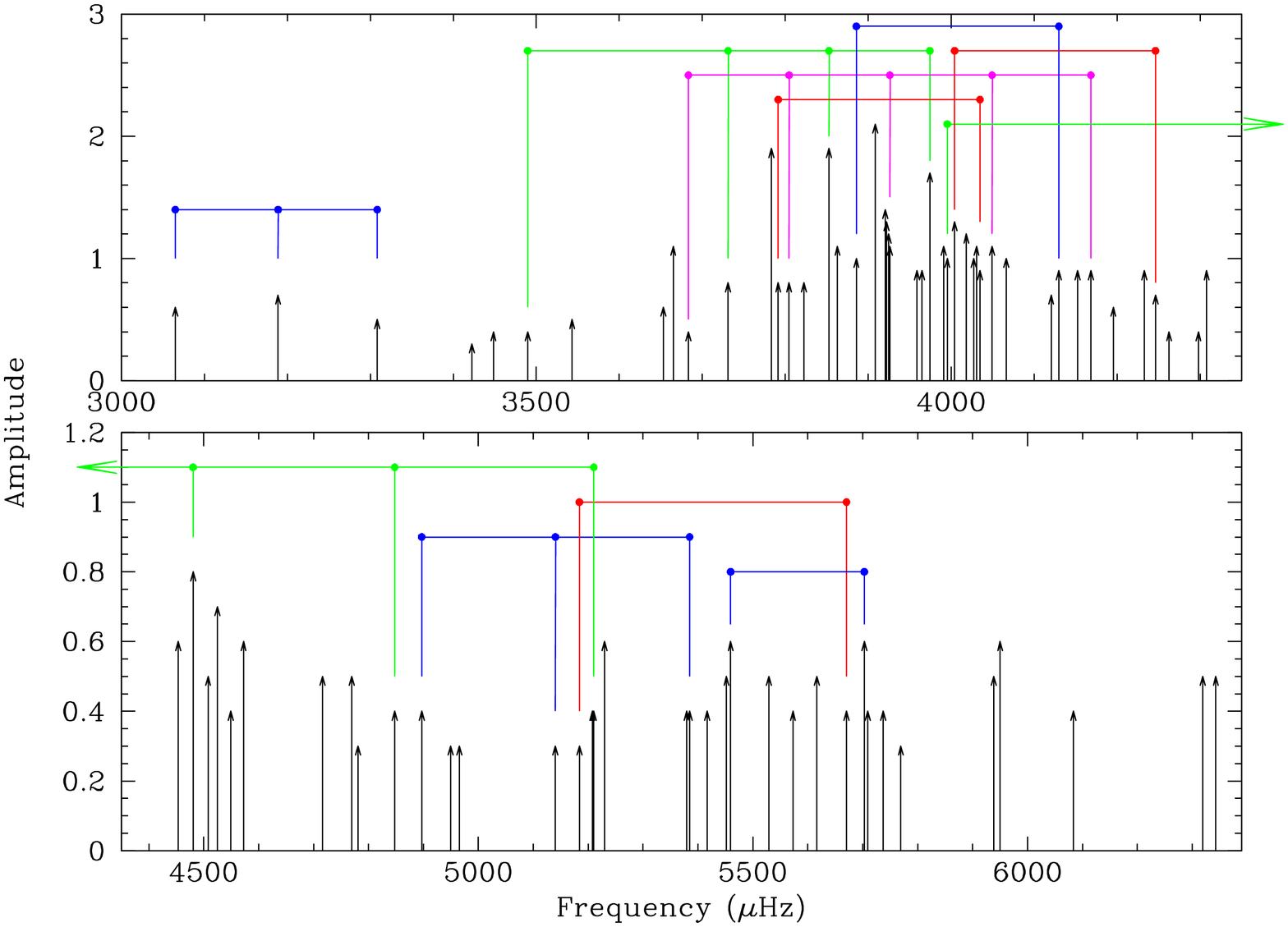,angle=-90,width=4.0in}
\epsfig{figure=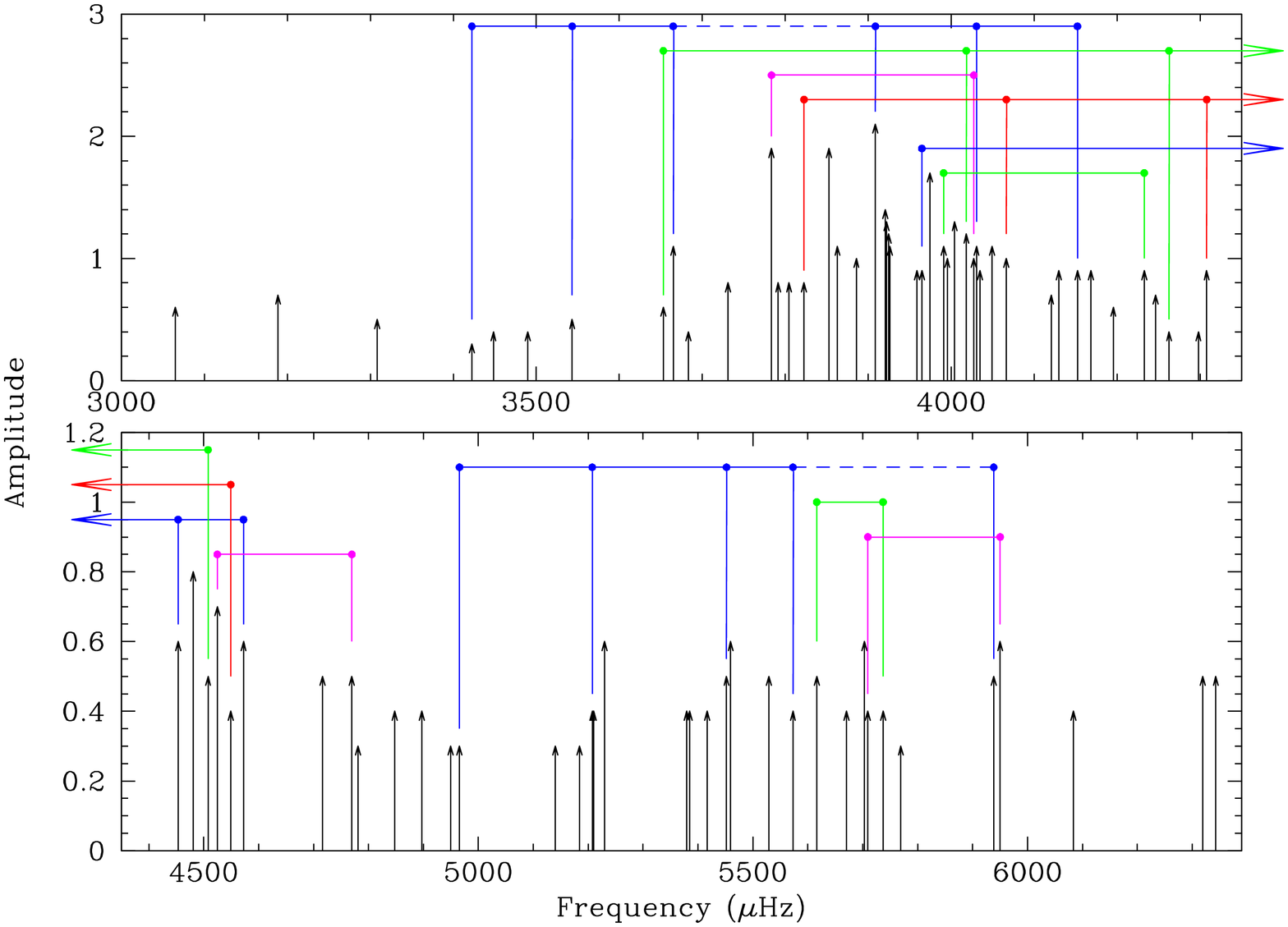,angle=-90,width=4.0in}
  \caption{A schematic of KPD1930's pulsation content to indicate
evenly spaced frequency multiplets (provided in
Table~8). For clarity,
each set of multiplets are connected via horizontal bars at differing
heights and have different colors. Dashed lines indicate possible,
but not probable multiplet members.
\label{fig09} }
\end{figure}
\pagebreak

\begin{figure}
 \centerline{
\epsfig{figure=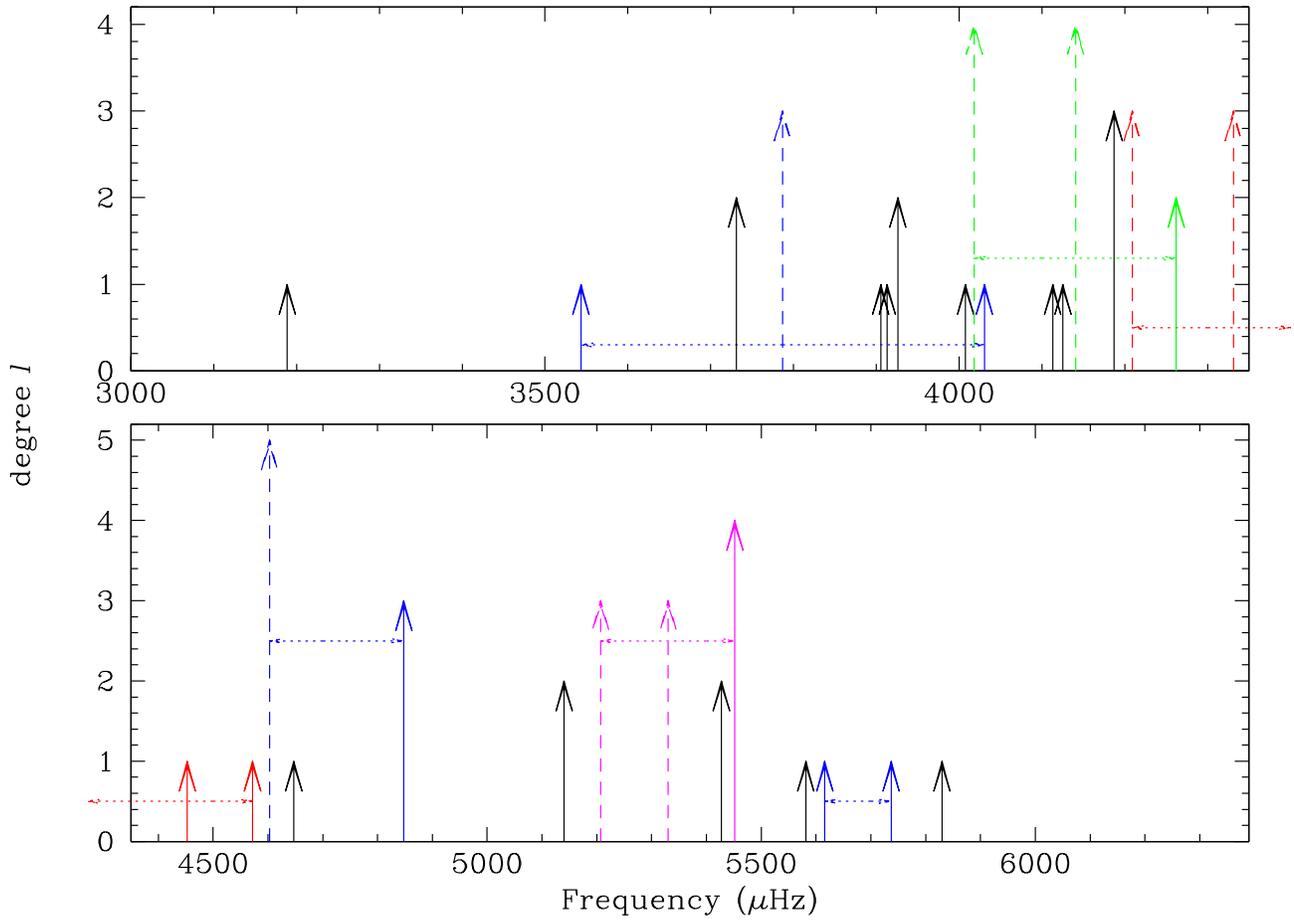,angle=-90,width=\textwidth}}
  \caption{Schematic showing the $m=0$ component of classically
interpreted multiplets.
The modal degree ($\ell$) is indicated by arrow
length. For those multiplets with more then one possible $m=0$ 
frequency, they are connected by a dotted horizontal line and for those
with more then one possible degree interpretation, the most-likely
value has a solid arrow while the less probable interpretation has a
dashed arrow. 
\label{fig10} }
\end{figure}
\pagebreak

\begin{figure}
 \centerline{
\epsfig{figure=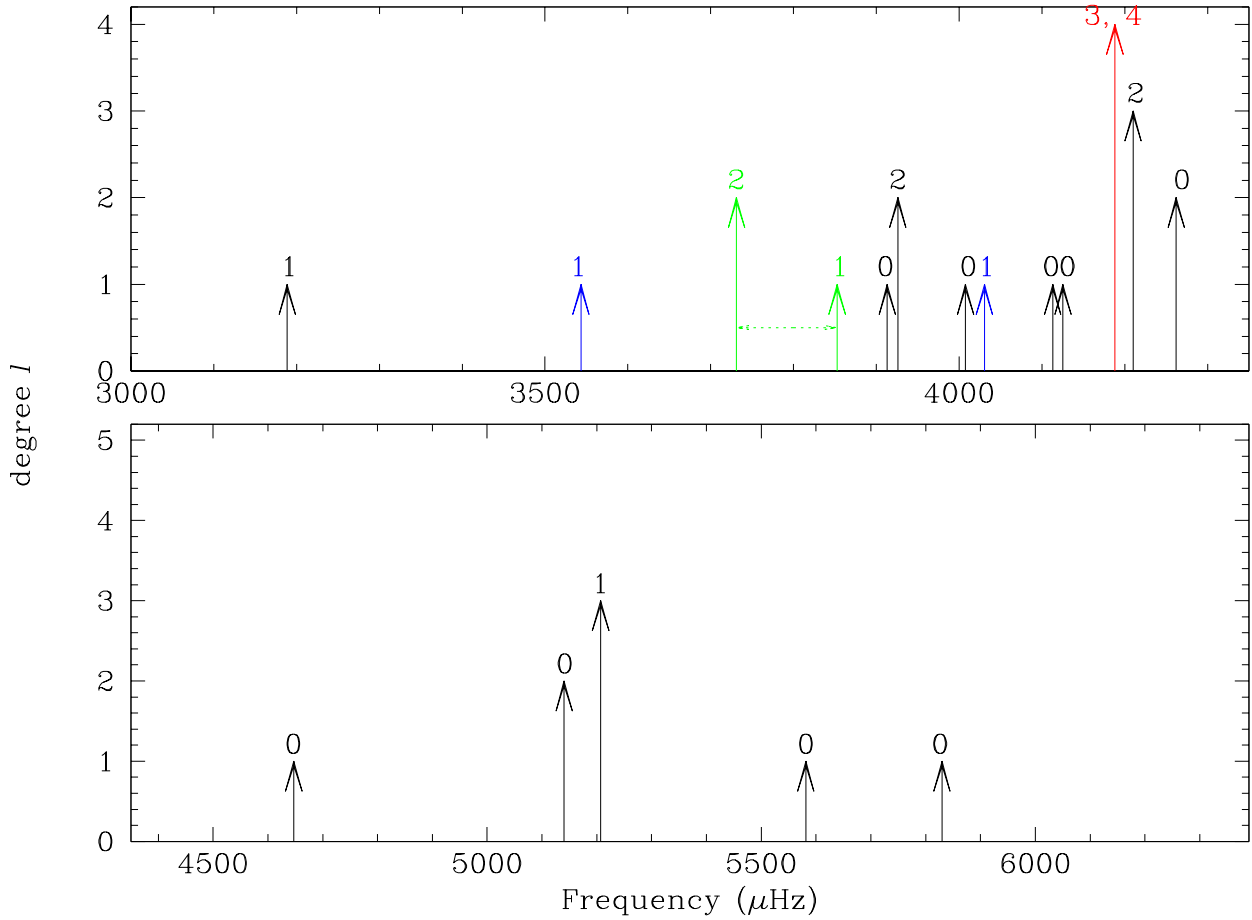,angle=-90,width=\textwidth}}
 \caption{Schematic showing possible frequencies and modes for a
'tipped' pulsation axis which points at the companion and precesses
with the orbit.
The modal degree ($\ell$) is indicated by arrow
length and the azimuthal order ($m$) by the number above each arrow.
For multiplets with more then one possible interpretation, the arrows
are connected by a dotted horizontal line and the most-likely
value (or combination) has a solid arrow while the less probable
interpretation has a dashed arrow. For clarity, modes with 
multiple interpretations
are color-coded. \label{fig15} }
\end{figure}
\pagebreak

\begin{figure}
 \centerline{
\epsfig{figure=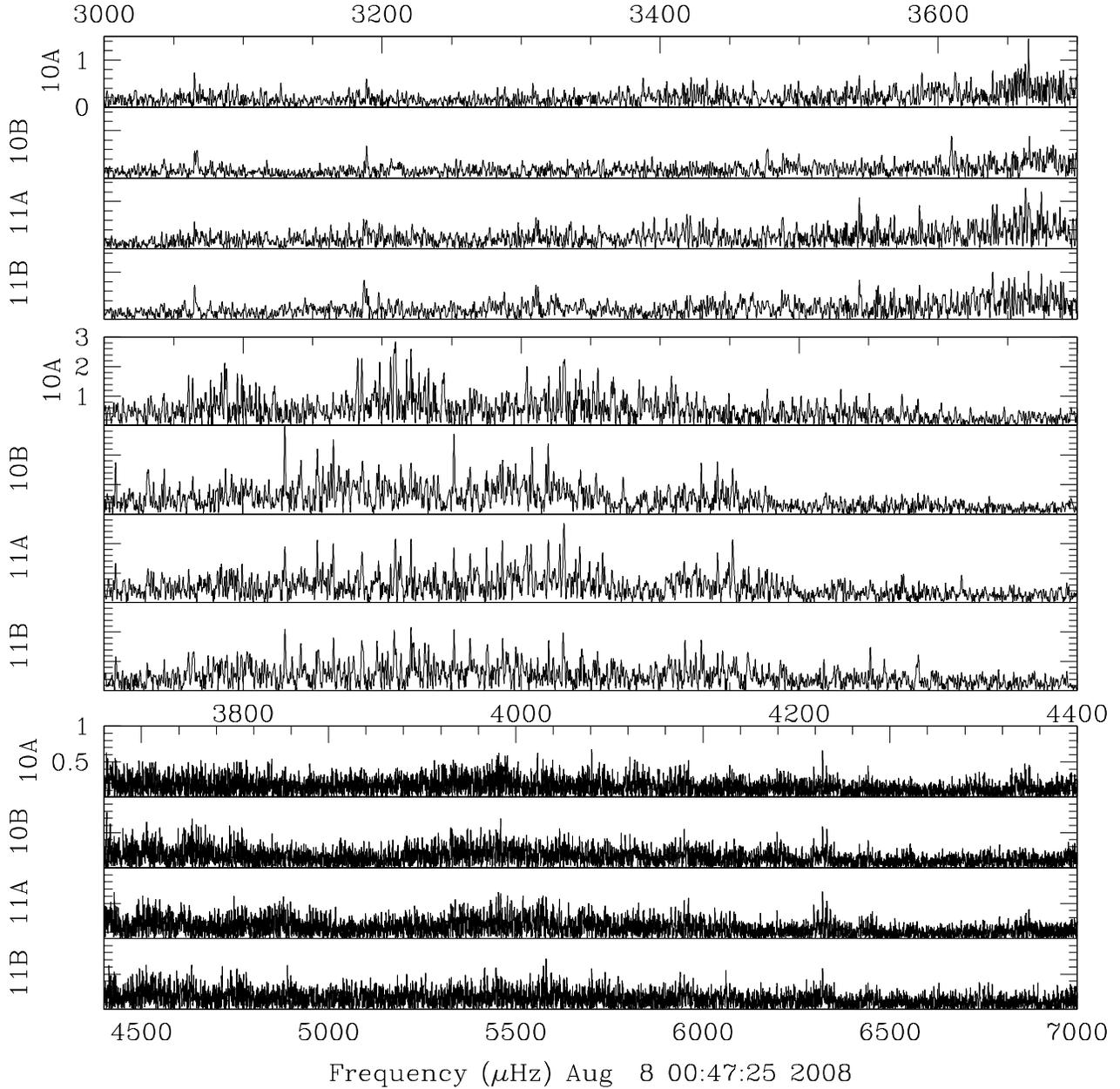,width=\textwidth}}
 \caption{Pulsation spectra of Group II's data, separated into
opposing phases (labeled A and B) appropriate for $\ell ,\,m=1,0$ 
and $1,1$ tipped pulsation axis modes. 
Data for each phase covers half of the orbital period
with set A of $\ell ,\,m\,=\,1,1$ including orbital phases from
0.0 to 0.5 and set B covering the other half. Those for 
$\ell ,\,m\,=\,1,0$ are shifted by $-0.25$ in phase. 
\label{fig16} }
\end{figure}